\documentclass[useAMS,usenatbib,usegraphicx]{mn2e}

\usepackage{threeparttable}

\newcommand{\scinum}[2]{$#1 \times 10^{#2}$}  

\newcommand{\scinuma}[3]{$(#1 \pm #2) \times 10^{#3}$}

\newcommand{\decnuma}[2]{$#1 \pm #2$}

\newcommand{\snxs}{\sigma_{\mathrm{NXS}}^{2}}

\newcommand{\nuhfb}{\nu_{\mathrm{HFB}}}

\newcommand{\nulfb}{\nu_{\mathrm{LFB}}}

\newcommand{\chfb}{C_{\mathrm{HFB}}}

\newcommand{\clfb}{C_{\mathrm{LFB}}}

\newcommand{\psdamp}{PSD_{\mathrm{AMP}}}

\newcommand{\nunyq}{\nu_{\mathrm{Nyq}}}

\newcommand{\mbh}{M_{\bullet}}

\newcommand{\msun}{\mathrm{M}_{\odot}}

\newcommand{\hzmdot}{\mathrm{Hz}~\mathrm{M}_{\odot}}

\newcommand{\hbeta}{FWHM(\mathrm{H}\beta)}

\title[X-ray  variability  amplitude and  black  hole  mass in  active
galactic nuclei]{The relationship  between X-ray variability amplitude
and black hole mass in active galactic nuclei}

\author[P.   M.  O'Neill,  K.  Nandra,  I.  E.   Papadakis and  T.  J.
Turner]{Paul  M.  O'Neill$^{1}$\thanks{E-mail: p.oneill@imperial.ac.uk
(PMO);   k.nandra@imperial.ac.uk   (KN);  jhep@physics.uoc.gr   (IEP);
turner@lucretia.gsfc.nasa.gov  (TJT)},   Kirpal  Nandra$^{1}$,  Iossif
E. Papadakis$^{2,3}$ and  T. Jane Turner$^{4,5}$ \\ $^{1}$Astrophysics
Group,  Imperial College London,  Blackett Laboratory,  Prince Consort
Road,  London  SW7~2BW\\ $^{2}$Department  of  Physics, University  of
Crete, 71 003, Heraklion, Crete, Greece\\ $^{3}$IESL, FORTH-Hellas, 71
110,  Heraklion,  Crete,  Greece\\  $^{4}$Laboratory for  High  Energy
Astrophysics, Code  660, NASA Goddard Space  Flight Center, Greenbelt,
MD 20771,  USA\\ $^{5}$University of Maryland,  Baltimore County, 1000
Hilltop Circle, Baltimore, MD 21250, USA\\}

\begin{document}

\date{Accepted. Received.}

\pagerange{\pageref{firstpage}--\pageref{lastpage}} \pubyear{2004}

\maketitle

\label{firstpage}

\begin{abstract}

We have  investigated the  relationship between the  X-ray variability
amplitude and  black hole mass for  a sample of  46 radio-quiet active
galactic nuclei  observed by \emph{ASCA}. Thirty-three  of the objects
in our  sample exhibited significant variability over  a time-scale of
$\sim$40~ks.   We determined  the  normalised excess  variance in  the
2--10~keV  light  curves of  these  objects  and  found a  significant
anti-correlation between excess variance  and black hole mass.  Unlike
most previous studies, we have quantified the variability using nearly
the  same  time-scale  for   all  objects.   Moreover,  we  provide  a
prescription  for  estimating  the  uncertainties  in  variance  which
accounts  both for  measurement uncertainties  and for  the stochastic
nature of  the variability.  We  also present an analytical  method to
predict the excess variance from a model power spectrum accounting for
binning, sampling and windowing  effects.  Using this, we modelled the
variance--mass  relation   assuming  all  objects   have  a  universal
twice-broken  power spectrum, with  the position  of the  breaks being
dependent  on  mass.   This  accounts  for the  general  form  of  the
variance--mass relationship  but is formally  a poor fit and  there is
considerable scatter.   We investigated this scatter as  a function of
the  X-ray  photon  index,  luminosity  and  Eddington  ratio.   After
accounting for the  primary dependence of excess variance  on mass, we
find  no  significant  correlation  with either  luminosity  or  X-ray
spectral slope.  We do  find an \emph{anti-correlation} between excess
variance and the  Eddington ratio, although this relation  might be an
artifact  owing  to  the   uncertainties  in  the  mass  measurements.
It   remains  to   be  established  that   enhanced  X-ray
variability is a property of  objects with steep X-ray slopes or large
Eddington ratios.  Narrow-line Seyfert  1 galaxies, in particular, are
consistent with being more variable than their broad line counterparts
solely because they tend to have smaller masses.
  
\end{abstract}

\begin{keywords}

galaxies:active    --    galaxies:nuclei    --   X-rays:galaxies    --
galaxies:Seyfert

\end{keywords}

\section{Introduction \label{sec:int}}

Variability was discovered in  the X-ray emission from active galactic
nuclei (AGNs)  roughly three decades  ago \citep[e.g.,][and references
therein]{mwp81}.   \emph{EXOSAT}  subsequently  obtained  well-sampled
light curves on time-scales of  minutes to days, and the power spectra
generated  from  these light  curves  were  described  as a  power-law
$P\propto  \nu^{-\alpha}$ with  a steep  `red-noise' index  of $\alpha
\sim 1.5$  and an amplitude  inversely proportional to  the luminosity
\citep{lp93,gml93}.  It  was clear that  this power-law must  break at
some lower  frequency, or the  power would diverge, and  some evidence
for   this   was  found   using   longer-term  archival   observations
\citep{m88,pm95}.   It  was  not  until  the  launch  of  \emph{RXTE},
however, that this break was measured definitively \citep{en99}.

A  number  of high  quality  power  spectra  have now  been  obtained,
primarily    using     \emph{RXTE}    and    \emph{XMM-Newton}    data
\citep[e.g.,][]{ump02,mev03,vfn03,mpu04,um04}.    These   have   shown
breaks to be common and emphasized the similarity of AGN power spectra
to that of the black hole  binary Cyg~X-1.  In the low/hard state, the
power  spectrum of  Cyg~X-1  exhibits a  twice-broken power-law  which
breaks from  a slope of $\alpha  \sim 0$ to $1$  at the `low-frequency
break'  ($\nulfb$)   and  from   $\alpha  \sim  1$   to  $2$   at  the
`high-frequency   break'   ($\nuhfb$),   with   $\nuhfb   \sim$1--6~Hz
\citep[e.g.,][]{bh90}.   In the  high/soft state,  the  power spectrum
exhibits  only  a  high-frequency  break,  with  $\nuhfb\sim$10--15~Hz
\citep[e.g.,][]{chr97,rgc00}.  Though  still a subject  of debate, the
emerging consensus is that we  usually see the high-frequency break in
the AGN power-spectra, although two  breaks are apparently seen in two
objects   \citep[viz,  AKN~564   and   NGC~3783;][]{pbn02,mev03}.   In
another,  the  narrow-line  Seyfert~1  (NLS1) NGC~4051,  there  is  no
low-frequency  turnover to  $\alpha=0$ down  to very  low frequencies,
which led \citet{mpu04} to  hypothesise that this object (and possibly
all NLS1s) resembled Cyg X-1 in the high/soft state.

Determining  accurate  power-spectra  for  AGN  is  difficult,  as  it
requires  high quality data  with near-even  sampling.  Such  data are
available only for a limited number  of objects and are very costly to
obtain in terms  of observing time.  It is  nonetheless very useful to
quantify  the   X-ray  variability  of  AGN  to   compare  with  other
properties,  and normalised  excess variance,  denoted as  $\snxs$, is
much simpler to calculate \citep{ngm97}. An anti-correlation was found
between excess variance  and luminosity for a sample  of AGNs observed
by  the  \emph{Advanced  Satellite  for  Cosmology  and  Astrophysics}
(\emph{ASCA}), confirming the \emph{EXOSAT}  results but with a larger
sample of  objects \citep{ngm97}.   Later work also  using \emph{ASCA}
data revealed that, for a  given luminosity, the X-ray light curves of
NLS1s exhibit  a larger excess  variance than the  classical Seyfert~1
galaxies \citep{tgn99,l99}.

\citet{ly01} and  \citet{bz03}, again using  \emph{ASCA} data, studied
the  relationship between  the  excess variance  (on  a time-scale  of
roughly  1~d) and  the black  hole  mass.  Those  studies revealed  an
anti-correlation between  $\snxs$ and  mass, which is  suggestive that
this  is the  primary relationship  rather than  with  luminosity. The
NLS1s appeared to follow the same relationship as the other AGN.

\citet{p04} investigated the  relationship between excess variance and
black  hole mass  on much  longer  time-scales (  $\sim 300$~d)  using
\emph{RXTE}  data on  a sample  of 10~AGNs.   The  classical Seyfert~1
galaxies followed a variance--mass  relation that is consistent with a
universal  power-spectral shape  as described  above for  the low/hard
state  of  Cyg~X-1.   In  the  universal model  used  by  \citet{p04},
$\nuhfb$  is  inversely  proportional  to  black hole  mass,  and  the
amplitude,  when represented  in  power $\times$  frequency space,  is
assumed to be constant.  In agreement with the power spectrum analysis
of McHardy  (2004), \citet{p04} found  that the NLS1 NGC~4051  did not
follow the same variance--mass relationship described by the classical
Seyfert 1s.   The excess  variance of NGC~4051  was consistent  with a
singly-broken power-law,  breaking from  $\alpha = 1$  to $2$,  with a
break-frequency $20$ times higher than  that deduced for the other the
Seyfert~1s.

These works  show that excess variance  can be a  useful complement to
full-blown power-spectral  analysis, and have the  advantage that they
can be  applied to a larger  number, and wider variety  of objects. As
has been  shown by \citet[][]{vew03},  some caution must  be exercised
when interpreting  excess variance measurements, primarily  due to the
red-noise shape of the power  spectra and the stochastic nature of the
variability. Such effects have not  been accounted for in the majority
of  previous works. The  intention of  the work  presented here  is to
investigate  the relationship between  excess variance  and mass  in a
large  sample of  AGN, improving  on these  previous studies  by fully
accounting  for  measurement  uncertainties,  sampling  and  red-noise
effects in the calculation of the excess variance and its uncertainty.

%%%%%%%%%%%%%%%%%%%%%%%%%%%%%%%%%%%%%%%%%%%%%%%%%%%%%%%%%%%%%%%%%%%%%%%%
%%%%%%%%%%%%%%%%%%%%%%%%%%%%%%%%%%%%%%%%%%%%%%%%%%%%%%%%%%%%%%%%%%%%%%%%

\begin{table*}

\begin{threeparttable}

\caption{X-ray spectral and variability information for objects having
at least  1 valid light  curve segment.  The 2--10~keV  luminosity and
hard-X-ray photon index are given for objects in which variability was
detected.}

\begin{tabular}{lccccccccl}
\hline

Name & $\mbh$ & $L_{\mathrm{X}}$ &  $\Gamma$ & Num. & Num. & $\snxs$ &
log $\snxs$ & $\Delta$log $\snxs$ & Refs. \\

& & &  & Seq.  & Seg. &  $\pm$ Boot. Unc. & $\pm$ Total  Unc.  & $\pm$
Total Unc. & \\

(1) & (2) & (3) & (4) & (5) & (6) & (7) & (8) & (9) & (10) \\

\hline

MRK~335 &  7.15 & 43.07  & 1.87 &  1 & 1 &  \scinuma{3.12}{1.87}{-3} &
\decnuma{-2.51}{0.41} & \decnuma{-0.44}{0.41} & R,1,2 \\

PG~0026$+$129 & 8.59 & 44.53 & 1.96 & 1 & 3 & \scinuma{1.31}{1.92}{-3}
& \decnuma{-2.88}{0.66} & \decnuma{0.61}{0.66} & R,1 \\

TON~S180 & 7.09  & 43.58 & 2.43 & 2 &  26 & \scinuma{1.59}{0.10}{-2} &
\decnuma{-1.80}{0.07} & \decnuma{0.21}{0.07} & L,3,4 \\

I~Zw~1 &  7.20 &  43.35 & 2.40  & 1  & 1 &  \scinuma{1.88}{0.92}{-2} &
\decnuma{-1.73}{0.39} & \decnuma{0.39}{0.39} & L,3,4 \\

F~9  & 8.41  &  43.91 &  1.91 &  8  & 6  & \scinuma{3.49}{5.52}{-4}  &
\decnuma{-3.46}{0.70} & \decnuma{-0.16}{0.70} & R,1,2 \\

RX~J0152.4$-$2319 & 7.87 & ... & ...  & 1 & 2 & $<$ \scinum{6.5}{-3} &
$<$ $-1.94$ & ... & L,5 \\

MRK~0586 &  7.86 & 44.07 & 2.22  & 1 & 3  & \scinuma{2.57}{0.75}{-2} &
\decnuma{-1.59}{0.22} & \decnuma{1.17}{0.22} & L,6,4 \\

MRK~1040 &7.64  & 42.40 &  1.69 & 1  & 1 &  \scinuma{1.20}{0.65}{-2} &
\decnuma{-1.92}{0.40} & \decnuma{0.62}{0.40} & S,6,7 \\

NGC~985 &  8.05 & 43.50  & 1.73 &  1 & 2 &  \scinuma{3.47}{1.76}{-3} &
\decnuma{-2.46}{0.32} & \decnuma{0.48}{0.32} & L,5 \\

1H~0419$-$577 & 8.58 & ... & ... & 2 & 3 & $<$ \scinum{4.31}{-3} & $<$
$-2.12$ & ... & L,3\\

F~303  & 6.37 &  43.03 &  1.92 &  1 &  1 &  \scinuma{6.72}{6.03}{-3} &
\decnuma{-2.17}{0.44} & \decnuma{-0.74}{0.44} & L,5 \\

AKN~120 &  8.18 & 43.88  & 1.93 &  1 & 2 &  \scinuma{3.78}{7.67}{-4} &
\decnuma{-3.42}{0.91} & \decnuma{-0.35}{0.91} & R,1,8 \\

PG~0804$+$761 & 8.84 & ... & ... & 1 & 2 & $<$ \scinum{3.37}{-3} & $<$
$-2.23$ & ... & R,1 \\

PG~0844$+$349 & 7.97 & ... & ... & 1 & 2 & $<$ \scinum{1.17}{-2} & $<$
$-1.69$ & ... & R,1 \\

MRK~110 &  7.40 & ...  &  ...  & 1 &  1 & $<$  \scinum{1.63}{-3} & $<$
$-2.55$ & ... & R,1 \\

PG~0953$+$415 & 8.44 & ... & ... & 1 & 2 & $<$ \scinum{8.18}{-3} & $<$
$-1.85$ & ...  & R,1 \\

NGC~3227 &  7.63 & 41.66 & 1.52  & 2 & 4  & \scinuma{2.41}{0.20}{-2} &
\decnuma{-1.62}{0.16} & \decnuma{0.91}{0.16} & R,1,2 \\

MRK~142 &  6.76 & 43.17  & 2.12 &  2 & 1 &  \scinuma{4.54}{1.33}{-2} &
\decnuma{-1.34}{0.34} & \decnuma{0.37}{0.34} & L,5,4 \\

HE~1029$-$1401   &   9.08    &   44.44   &   1.83   &    1   &   2   &
\scinuma{1.02}{1.21}{-3}        &        \decnuma{-2.99}{0.56}       &
\decnuma{0.98}{0.56} & L,6,9 \\

NGC~3516 & 7.63  & 43.08 & 1.83 & 5 &  18 & \scinuma{3.70}{0.45}{-3} &
\decnuma{-2.43}{0.10} & \decnuma{0.10}{0.10} & R,1,2 \\

PG~1116$+$215 & 8.21 & ... & ... & 1 & 1 & $<$ \scinum{1.06}{-2} & $<$
$-1.73$ & ... & L,6 \\

EXO~1128.1+6908 & 7.02 &  ... & ... & 1 & 1  & $<$ \scinum{1.78}{-2} &
$<$ $-1.51$ & ... & L,5 \\

NGC~3783 &  7.47 & 42.90 & 1.70  & 9 & 8  & \scinuma{3.91}{0.51}{-3} &
\decnuma{-2.41}{0.13} & \decnuma{-0.03}{0.13} & R,1,2 \\

NGC~4051 &  6.28 & 41.21 & 1.92  & 2 & 6  & \scinuma{8.62}{0.66}{-2} &
\decnuma{-1.06}{0.09} & \decnuma{0.31}{0.09} & R,1,2\\

NGC~4151 & 7.12 & 42.62 &  1.53 & 13 & 29 & \scinuma{2.79}{0.22}{-3} &
\decnuma{-2.55}{0.07} & \decnuma{-0.51}{0.07} & R,1,2 \\

PG~1211$+$143 & 8.16 & ... & ... & 1 & 1 & $<$ \scinum{2.39}{-2} & $<$
$-1.38$ & ... & R,1 \\

MRK~766 &  6.54 & 42.73  & 2.16 &  1 & 2 &  \scinuma{4.02}{0.48}{-2} &
\decnuma{-1.40}{0.16} & \decnuma{0.15}{0.16} & L,6,2 \\

NGC~4395 &  4.11 & 39.99  & 1.7 &  5 & 6 &  \scinuma{1.13}{0.14}{-1} &
\decnuma{-0.95}{0.10} & \decnuma{0.17}{0.10} & L,10,11 \\

NGC~4593 &  6.73 & 42.98 & 1.81  & 2 & 1  & \scinuma{1.42}{0.21}{-2} &
\decnuma{-1.85}{0.33} & \decnuma{-0.16}{0.33} & R,1,8 \\

WAS~61 &  6.66 & ...   & ...  &  1 & 1  & $<$ \scinum{6.95}{-3}  & $<$
$-1.92$ & ... & L,5 \\

PG~1244+026 & 6.07 & 43.03 & 2.46 & 1 & 2 & \scinuma{2.60}{0.62}{-2} &
\decnuma{-1.59}{0.18} & \decnuma{-0.31}{0.18} & L,5,12\\

MCG$-$6-30-15   &   6.19    &   42.72   &   2.00   &    6   &   48   &
\scinuma{4.16}{0.13}{-2}        &        \decnuma{-1.38}{0.03}       &
\decnuma{-0.05}{0.03} & L,3,2 \\

IC~4329A &  7.00 & 43.59 & 1.71  & 5 & 6  & \scinuma{2.36}{2.44}{-4} &
\decnuma{-3.63}{0.47} & \decnuma{-1.70}{0.47} & R,1,2 \\

MRK~279 &  7.54 & 43.66  & 1.99 &  1 & 1 &  \scinuma{2.32}{0.84}{-3} &
\decnuma{-2.63}{0.36} & \decnuma{-0.19}{0.36} & R,1 \\

NGC~5506 &  7.94 & 42.73 & 2.08  & 1 & 2  & \scinuma{1.06}{0.14}{-2} &
\decnuma{-1.97}{0.23} & \decnuma{0.87}{0.23} & S,13,8 \\

NGC~5548 & 7.83 & 43.41 &  1.79 & 11 & 16 & \scinuma{9.42}{2.67}{-4} &
\decnuma{-3.03}{0.14} & \decnuma{-0.30}{0.14} & R,1,2\\

MRK~1383 &  9.11 & ...  &  ... & 1 &  1 & $<$  \scinum{6.33}{-3} & $<$
$-1.96$ & ... & R,1 \\

MRK~478 &  7.34 & 43.50  & 2.06 &  1 & 2 &  \scinuma{6.14}{3.75}{-3} &
\decnuma{-2.21}{0.35} & \decnuma{0.04}{0.35} & L,3,4 \\

MRK~841 &  8.10 & 43.54  & 2.00 &  3 & 5 &  \scinuma{1.14}{0.93}{-3} &
\decnuma{-2.94}{0.38} & \decnuma{0.05}{0.38} & L,6,2 \\

MRK~290 &  7.05 & 43.22  & 1.77 &  1 & 2 &  \scinuma{4.11}{2.15}{-3} &
\decnuma{-2.39}{0.32} & \decnuma{-0.41}{0.32} & L,3,7 \\

IRAS~17020+4544   &   6.77   &   43.73    &   2.37   &   1   &   2   &
\scinuma{5.47}{2.00}{-3}        &        \decnuma{-2.26}{0.28}       &
\decnuma{-0.54}{0.28} & L,14,4 \\

MRK~509 &  8.16 & 44.03 & 1.82  & 11 & 2  & \scinuma{5.75}{7.17}{-4} &
\decnuma{-3.24}{0.59} & \decnuma{-0.19}{0.59} & R,1,2 \\

AKN~564 & 6.06  & 43.38 & 2.58 & 13 &  70 & \scinuma{5.34}{0.14}{-2} &
\decnuma{-1.27}{0.03} & \decnuma{0.00}{0.03} & L,3,4 \\

RX~J2248.6$-$5109 & 7.67 & ... & ... & 1 & 1 & $<$ \scinum{1.08}{-2} &
$<$ $-1.73$ & ... & L,5 \\

NGC~7469 &  7.09 & 43.25 & 1.84  & 3 & 2  & \scinuma{4.68}{1.60}{-3} &
\decnuma{-2.33}{0.27} & \decnuma{-0.32}{0.27} & R,1,2 \\

MCG$-$2-58-22 & 8.54 & ... & ... & 2 & 4 & $<$ \scinum{1.53}{-3} & $<$
$-2.58$ & ...  & L,3 \\

\hline

\end{tabular}

{\small The objects are listed in  order of R.A. (1) Object name.  (2)
Log  of black hole  mass in  units of  $\msun$. (3)  Log of  2--10 keV
luminosity  in units  of erg~s$^{-1}$.   (4) Hard-X-ray  photon index.
(5) Number of  available \emph{ASCA} observing  sequences.  (6) Number
of usable  light curve segments.  (7) Mean  normalised excess variance
with the uncertainty  or upper limit as determined  from the bootstrap
simulations.  (8) Log of the  mean normalised excess variance with the
uncertainty as  determined by combining the  bootstrap uncertainty and
the derived  red-noise scatter.   (9) Residuals from  the best-fitting
universal model  with the uncertainty  as determined by  combining the
bootstrap uncertainty and the  derived red-noise scatter.  (10) Method
used to determine the black hole  mass and references for the mass and
X-ray spectral  properties.  The methods, in order  of preference, are
as follows: R, reverberation  mapping; S, stellar velocity dispersion;
L,   relationship  between  broad-line   region  radius   and  optical
luminosity.   References:  1,   \cite{pfg04};  2,  \citet{ngm97b};  3,
\citet{bz03}; 4,  \citet{l99b}; 5, \citet{gwl04};  6, \citet{wu02}; 7,
\citet{r97}; 8,  \citet{np94}; 9, \citet{rt00};  10, \citet{fh03}; 11,
\citet{ifa00}; 12,  \citet{gty00}; 13, \citet{p04};  14, \citet{wl01}.
The reference for the black hole mass of each object is listed first.}

\label{tab:geninfo}

\end{threeparttable}
\end{table*}

%%%%%%%%%%%%%%%%%%%%%%%%%%%%%%%%%%%%%%%%%%%%%%%%%%%%%%%%%%%%%%%%%%%%%%%%%%%
%%%%%%%%%%%%%%%%%%%%%%%%%%%%%%%%%%%%%%%%%%%%%%%%%%%%%%%%%%%%%%%%%%%%%%%%%%%

\section{The Tartarus database and the AGN sample}

The          Tartarus\footnote{http://astro.ic.ac.uk/Research/Tartarus}
database contains  products for \emph{ASCA}  observations with targets
designated as AGN \citep{tnt01}.  We selected radio-quiet objects that
have data in  the Tartarus (Version 3) database and  also for which we
could  conveniently  obtain a  measurement  of  the  black hole  mass,
$\mbh$.   Seyfert~2 objects were  excluded from  our sample,  with the
exception  of NGC~5506  because for  this object  we are  confident of
seeing the X-ray emission  directly \citep{bww90}. This initial sample
comprised  68 AGN.   We  utilised the  Tartarus  analysis pipeline  to
extract light curves for the objects in this sample. As we describe in
detail in  the following Section,  not all light curves  were suitable
for our analysis.  Having  screened the available data, there remained
46  objects  for  which  we  could  suitably  characterise  the  X-ray
variability.   These objects  are  listed in  Table~\ref{tab:geninfo}.
Note that, while a flux limit  was not formally applied to our sample,
the effect  of the screening process  was to exclude  objects having a
low counting rate.

Recent progess  in measuring black  hole masses has made  possible the
work    we    present    here.     We    preferentially    used    the
reverberation-mapping mass  estimate from \citet{pfg04}.   If this was
not available then we used the mass estimate as determined from either
the  stellar  velocity   dispersion  \citep{gbb00}  or  the  empirical
relationship  between  the  broad-line  region radius  and  5100~$\AA$
luminosity    \citep{wpm99}.      The    masses    are     given    in
Table~\ref{tab:geninfo}  where  we  also   list  the  method  used  to
determine  the mass and  the corresponding  reference. The  masses for
most  objects were  available in  the  literature.  For  8 objects  in
Table~\ref{tab:geninfo} we obtained  optical spectral information from
\citet{gwl04} and utilised eqn.~6 from \citet{ksn00} and eqns.~1 and 2
from \citet{wu02} to determine $\mbh$.

The   2--10~keV  luminosity  $L_{2-10~\mathrm{keV}}$   and  hard-X-ray
(either 2--10~keV or 3--10~keV)  photon index $\Gamma$ are also listed
in  Table~\ref{tab:geninfo} for  those  objects in  which we  detected
variability.   The majority  of  $L_{2-10~\mathrm{keV}}$ and  $\Gamma$
values  were  taken  from \citet{ngm97b},  \citet{l99b},  \citet{r97},
\citet{np94}, \citet{rt00}, \citet{ifa00}, and \citet{gty00}.  For the
objects  PG~0026+129,  NGC~985,  F~303,  and MRK~279,  we  fitted  the
available  \emph{ASCA}  data  to  obtain  $L_{2-10~\mathrm{keV}}$  and
$\Gamma$.   The  SIS0,  SIS1,  GIS2,  and  GIS3  spectra  were  fitted
simultaneously in  the 2--10~keV rest-frame energy range.   We used an
absorbed  power-law, with $N_{\mathrm{H}}$  constrained to  be greater
the  galactic value  which we  obtained  using the  NASA HEASARC  `nH'
tool.\footnote{http://heasarc.gsfc.nasa.gov/cgi-bin/Tools/w3nh/w3nh.pl}
For  our  luminosity  calculations  we  obtained  redshifts  from  the
NASA/IPAC                                                 Extragalactic
Database\footnote{http://nedwww.ipac.caltech.edu/}  and used  $H_{0} =
75$~km~s$^{-1}$  and $q_{0}=0.5$.  All  $L_{2-10~\mathrm{keV}}$ values
collected from  the literature were  transformed to this  cosmology as
required.

\section{Excess variance analysis}

The  number  of \emph{ASCA}  observing  sequences  available for  each
object is  shown in Table~\ref{tab:geninfo}. We  extracted a 2--10~keV
combined  SIS0$+$SIS1$+$GIS2$+$GIS3 light  curve  from each  sequence.
These initial light  curves had a resolution of 16~s  and each bin was
required to be fully exposed.   The light curves were then rebinned to
a resolution of 256~s.

\subsection{Excess variance calculation}

For a red-noise process, the variance in a light curve depends both on
the power spectrum  of the variations and also  on the time resolution
and duration  of the light  curve.  This means that  different $\snxs$
measurements  are only  strictly comparable  if the  durations  of the
light  curves are equal.   Therefore, we  sub-divided the  light curve
from  each  sequence into  many  segments  of  similar duration.   The
advantage of using long durations is that the amplitude of variability
increases, and the number of  points used to calculate $\snxs$ is also
larger,  reducing the  measurement  uncertainty.  On  the other  hand,
using short durations has the advantage that more light curve segments
can be  included. We chose a  nominal segment length of  40~ks for our
analysis  as a tradeoff  between these  considerations. In  reality we
chose a duration of 39936~ks which is an integer multiple of our 256~s
time bin.

The sub-dividing of the light  curves proceeded as follows. First, the
earliest  40~ks segment  of the  light curve  for a  certain observing
sequence  was selected.   Then, beginning  with the  next  exposed bin
following the  end of  this first segment,  another 40~ks  segment was
selected.  This  continued until the  light curve had  been completely
sub-divided.   Note that the  actual duration  of these  light curves,
which we  define as the time  between the first and  last exposed bin,
can be less than 40~ks because the dividing point between segments can
occur when there is a gap in the data train. We accepted all resulting
light curve segments that had a duration $>$30~ks.

To ensure Gaussian  statistics, we required each 256~s  bin to contain
at  least 20~counts.   The number  of counts  in a  certain  256~s bin
depends both on  the source counting rate and  the fractional exposure
of  that bin.  We  do not  wish to  reject entire  observing sequences
simply because  some of  the bins  have a low  exposure, but  those in
which fully-exposed bins have $<20$  counts should be excluded.  If we
were to  remove bins simply because  they had a low  counting rate, we
would be biased against observing objects when their intensity is low.
Selecting  according to fractional  exposure, on  the other  hand, can
remove bins  having too few  counts without introducing this  bias, as
fractional exposure  is not  related to the  intensity of  the source.
If, for example,  there are fully exposed bins  with $<20$ counts, the
entire sequence  was discarded. This is  the case when  we are dealing
with a weak source.  For brighter sources where only underexposed bins
have $<20$  counts, we excise  all bins below some  minimal fractional
exposure.  This gets rid of the non-Gaussian bins, allowing us to keep
the remainder of the light  curve for further analysis, but introduces
no bias against  those times when the source is weak  due to true flux
variability.

Finally, we  further required the  truncated and screened  light curve
segments  to have  at least  20 bins,  so that  the variance  could be
determined accurately.

This procedure  resulted in 46 objects  having at least  1 valid light
curve segment, and  a total of 305 valid segments  in all.  The number
of segments for each  object is given in Table~\ref{tab:geninfo}.  The
mean durations of the light curve segments for each object were in the
range 35--40~ks in the observers  frame.  The 48 objects in our sample
have  redshifts in the  range 0.001--0.234.   Taking into  account the
redshift of  each object,  the rest-frame mean  durations were  in the
range  30--40~ks.  We expect  the effect  of these  slightly different
durations to be small.  For a power spectrum with a power-law slope of
$\alpha = 2$, the worst  case we expect, a $\sim$25~per~cent reduction
in the light  curve duration (i.e., from 40~ks to  30~ks) results in a
reduction in the $\snxs$ of only $\sim$0.1~dex.  As presented later in
this Section, the uncertainties in most of our observed $\snxs$ values
are  a few  to  several  times larger  than  0.1~dex.  Therefore,  the
$\sim$25~per~cent  difference between  the shortest  and  longest mean
light curve duration  can be neglected and allows us  to use more data
than would  have been available  if we had  imposed a strict  limit on
duration.

We tested  to see which  objects exhibited significant  variability by
performing  a chi-squared  test. The  $\chi^{2}$ corresponding  to the
hypothesis  of  a  constant  counting  rate was  determined  for  each
$\sim$40~ks light curve segment.  Then, for each object, we summed all
of the $\chi^{2}$s and  degrees-of-freedom (DOFs) to test whether that
object  is variable.   We detected  variability in  33 objects  at the
95~per~cent confidence level.  We  then calculated the excess variance
in each light curve segment with the following expression:

\begin{equation}
\snxs  =  \frac{1}{N  \mu_2}   \sum^{N}_{i=1}  [(X_{i}  -  \mu)^{2}  -
\sigma_{i}^{2} ]
\end{equation}

\noindent where $N$ is the number  of bins in the segment, $X_{i}$ and
$\sigma_{i}$ are  the counting rates  and uncertainties, respectively,
in  each bin,  and  $\mu$ is  the  unweighted arithmetic  mean of  the
counting rates.   For objects  with more than  one valid  segment, the
unweighted average excess variance  was determined.  A major advantage
of our work is that, given the large number of light curves available,
there  is  often   more  than  one  valid  segment   per  object  (see
Table~\ref{tab:geninfo}).  Taking  the mean $\snxs$  of these multiple
segments  reduces  the  potentially  large uncertainty  owing  to  the
stochastic nature of the variability (see below). When calculating the
mean excess variance we used \emph{all} valid light curve segments for
a  particular  object,  including  those  segments that  did  not,  in
themselves, exhibit variability based on the $\chi^{2}$ test.

\subsection{Estimating the uncertainties in $\snxs$}

Estimating   the   uncertainty  for   excess   variance  is   somewhat
complicated.   Analytical   prescriptions  have  been   given  in  the
literature by \citet[][their correct formula is given by Turner et~al.
1999]{ngm97},  \citet{etp02} and  \citet{vew03}.   The latter  authors
also discussed the uncertainties in  $\snxs$ on the basis of simulated
red-noise light curves. These uncertainties depend both on measurement
uncertainties (e.g., Poisson  noise) in the light curve  data, and the
stochastic nature  of the variability:  any given light  curve segment
represents  just one  realisation of  a random  process, and  thus can
exhibit a different mean and  variance from the true value, or another
random  segment.    This  `noise'  uncertainty  can   be  very  large,
especially for a single realisation.  One must, therefore, account for
this uncertainty  before apparent differences in $\snxs$,  either in a
given     source    \citep{np01}     or    in     comparing    sources
\citep[e.g.,][]{tgn99} can be  considered robust.  The measurement and
noise  uncertainties on  $\snxs$ are  unrelated,  so can  and must  be
estimated separately.

To  estimate   the  uncertainty   in  $\snxs$  owing   to  measurement
uncertainties,  we  used bootstrap  simulations  \citep[the reader  is
directed  to][for  a   discussion  on  bootstrap  simulations]{ptv01}.
Suppose that the  observed light curve contains $N$  bins.  This light
curve  is  a distribution  of  $N$  counting  rates and  corresponding
Poisson-noise uncertainties from which we calculate $\snxs$. Note that
calculating $\snxs$ does not depend on the bins being in time-order. A
bootstrap   simulation   involves   randomly  selecting,   from   that
distribution,  a new  set of  $N$ bins.   The duplication  of  bins is
permitted  during the selection  process. This,  then, results  in the
creation of  a slightly different distribution of  counting rates, and
$\snxs$ can  be determined for  this new distribution. If  one repeats
the entire process many times, the resulting distribution of simulated
$\snxs$ values provides an estimate of the uncertainty in $\snxs$.

We performed a series of  10000 bootstrap simulations to determine the
uncertainty in the \emph{mean} observed $\snxs$ for each object in our
sample. Each  of these simulations  involved: simulating a  new `light
curve' from  each valid light  curve segment, determining  $\snxs$ for
those simulated  light curves,  and then determing  the mean  of these
simulated  $\snxs$  values.   We  were  thus able  to  generate  10000
simulated values of the mean  $\snxs$. The standard deviation of these
values  was  taken to  be  the  measurement  uncertainty in  the  mean
observed  $\snxs$.    We  refer  to  this  value   as  the  `bootstrap
uncertainty' and denote it as $\Delta_{\mathrm{boot}}(\snxs)$.

Estimates  of the  uncertainty owing  to the  noise process  have been
presented  by \citet{vew03},  based  on light  curve simulations,  who
showed that the noise uncertainty is proportional to the mean value of
the variance.   The constant of  proportionality depends on  the power
spectrum shape, which  we do not know a  priori. Therefore, as pointed
out by \citet{vew03}, it  is preferable to determine the uncertainties
in $\snxs$ directly from the data.

Our large  database contains  6 objects (viz,  AKN~564, MCG$-$6-30-15,
TON~S180, NGC~4151,  NGC~3516, and NGC~5548) with  a sufficient number
of light curve  segments ($>$15) to make a  meaningful estimate of the
fractional uncertainty in  $\snxs$ owing to noise nature  of our light
curves.     First,    we    determined    the    standard    deviation
$\sigma_{\mathrm{obs}}$  of  the  observed  $\snxs$  values  for  each
object.   We  then determined,  from  the  bootstrap uncertainty,  the
standard  deviation $\sigma_{\mathrm{meas}}$ that  we would  expect to
observe in the  distribution of the $\snxs$ values  if the scatter was
owing    only   to    measurement   uncertainties.     We   subtracted
$\sigma_{\mathrm{meas}}$ from  $\sigma_{\mathrm{obs}}$, in quadrature,
to  obtain  the standard  deviation  $\sigma_{\mathrm{noise}}$ in  the
observed $\snxs$ values that is owing only to the stochastic nature of
the variability.  We then determined, for each of the 6 distributions,
the  ratio  between $\sigma_{\mathrm{noise}}$  and  mean $\snxs$.   We
shall refer to  this ratio as the `fractional  standard deviation' and
denote  it  as   $\sigma_{\mathrm{frac}}$.   The  fractional  standard
deviations   for   the   6   objects  were:   0.49   (AKN~564),   0.47
(MCG$-$6-30-15),  0.69 (TON~S180),  0.79 (NGC~4151),  0.82 (NGC~3516),
and  0.61 (NGC~5548).  These  values of  $\sigma_{\mathrm{frac}}$ show
that, \emph{even in the absence of measurement uncertainties}, one can
expect  noise  uncertainties in  the  range $\sim$50--80~per~cent  for
individual  measurments of  $\snxs$  \citep[see also][]{vew03}.   This
highlights  the  need  of  obtaining  many  realisations  (i.e.,  many
measurements of $\snxs$), regardless of  the level of Poisson noise in
the data.

\begin{figure}

  \rotatebox{270}{\includegraphics[height=1\columnwidth]{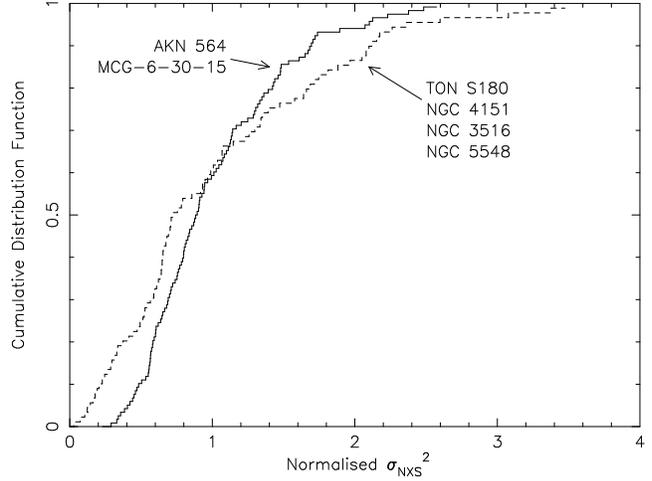}}

\caption{Cumulative distribution functions  of the combined normalised
$\snxs$ distributions  of AKN~564  and MCG$-$6-30-15 (solid  line) and
TON~S180, NGC~4151, NGC~3516, and NGC~5548 (dahed-line).}

\label{fig:snxs_cdf}
\end{figure}

\citet{vew03} showed that the uncertainty in the estimated variance of
a  red-noise light  curve increases  with the  steepness of  the power
spectrum slope.  Power-spectral analyses of AGN have revealed that the
value of $\nuhfb$ generally decreases with increasing black hole mass.
This means that the shape of the power spectrum in the frequency range
probed    by    our    light   curves    ($\sim$\scinum{2.5}{-5}    to
\scinum{4}{-3}~Hz) is expected  to vary as a function  of $\mbh$, such
that the objects  with the highest $\mbh$ should  exhibit the steepest
($\alpha \sim 2$) power spectra.  We expect, then, that the scatter in
$\snxs$  owing to  red-noise  fluctuations should  also increase  with
mass.  The lowest-mass objects  are AKN~564 and MCG$-$6-30-15, and the
observed values of $\nuhfb$ for  these fall within the frequency range
of    our    data    \citep{pbn02,vfn03}.    We    therefore    expect
$\sigma_{\mathrm{frac}}$ for this pair of  objects to be less than the
others.    This   does   indeed   appear   to   be   the   case:   the
$\sigma_{\mathrm{frac}}$ values of  AKN~564 and MCG$-$6-30-15 are both
less  than  those  of  TON~S180,  NGC~4151,  NGC~3516,  and  NGC~5548.
However,  we  possess only  a  limited  number  of individual  $\snxs$
measurements to estimate both the  mean and standard deviation of each
distribution, so it  is possible that this apparent  difference is not
statistically significant.   We decided,  therefore, to compare  the 6
distributions of  $\snxs$ values using a  series of Kolmogorov-Smirnov
(K-S) tests.  Before we could  compare the distributions, we first had
to correct each  of them for the effect  of measurement uncertainties.
To do this, we scaled the deviations of the observed $\snxs$ values so
that the standard deviation  of the `corrected' distribution was equal
to   $\sigma_{\mathrm{noise}}$,  with   the  mean   $\snxs$  remaining
unchanged.  We then normalised  each of the corrected distributions by
dividing the $\snxs$ values by the mean.  The corrected and normalised
$\snxs$  distributions for  AKN~564 and  MCG$-$6-30-15  are consistent
with being  drawn from the same  distribution.  The same  is true when
comparing the other 4 distributions  with each other.  We then created
two, combined  distributions: one  for AKN~564 and  MCG$-$6-30-15; and
another   for  TON~S180,  NGC~4151,   NGC~3516,  and   NGC~5548.   The
fractional standard  deviations from these  two combined distributions
were  0.48  (AKN~564,  MCG$-$6-30-15)  and 0.74  (TON~S180,  NGC~4151,
NGC~3516, NGC~5548), and a K-S test showed them to be different at the
95~per~cent confidence  level.  The cumulative  distribution functions
of     the     combined     distributions     are     presented     in
Fig.~\ref{fig:snxs_cdf}.    Combining    the   normalised,   corrected
distributions of all 6  objects resulted in a $\sigma_{\mathrm{frac}}$
of 0.61.

For the objects  AKN~564, MCG$-$6-30-15, TON~S180, NGC~4151, NGC~3516,
and     NGC~5548,    we     determined    the     total    uncertainty
[$\Delta_{\mathrm{tot}}(\snxs)$] in the  mean excess variance directly
from their respective values  of $\sigma_{\mathrm{obs}}$.  For each of
the other objects, we estimated  the noise uncertainty and combined it
in      quadrature       with      the      bootstrap      uncertainty
[$\Delta_{\mathrm{boot}}(\snxs)$] using the following expression:

\begin{equation}
\Delta_{\mathrm{tot}}(\snxs)      =      \sqrt{     \left(      \frac{
\sigma_{\mathrm{frac}}  \snxs}{\sqrt{N_{\mathrm{seg}}}}  \right)^{2} +
[\Delta_{\mathrm{boot}}(\snxs)]^{2} }
\end{equation}

\noindent   where   $\snxs$   is   the  mean   excess   variance   and
$N_{\mathrm{seg}}$   is   the   number   of  available   light   curve
segments. For objects  with log~$\mbh > 6.54$ we  adopted a fractional
standard deviation  of $\sigma_{\mathrm{frac}} = 0.74$,  while for the
objects   with  log~$\mbh   \leq   6.54$  we   adopted   a  value   of
$\sigma_{\mathrm{frac}} =  0.48$.  These ranges in  mass were selected
on the basis that the object  MRK~766, which has log~$\mbh = 6.54$, is
the most massive object that has an observed $\nuhfb$ in the frequency
ranged probed  by our  $\sim$40~ks light curves  \citep[e.g.,][and see
Introduction] {pbn02,vf03,mfm04,vif04}.  For objects more massive than
this we expect $\nuhfb$ to  be less than our observed frequency range.
In the absence of a measurement of $\mbh$ or any information regarding
the   shape    of   the   power   spectrum,   the    mean   value   of
$\sigma_{\mathrm{frac}} = 0.61$ can be adopted.

The  $\snxs$  upper limits  for  the  non-variable  objects were  also
estimated  by  combining  the   two  components  of  uncertainty.   We
multiplied  the  $1\sigma$ bootstrap  uncertainty  by the  appropriate
fractional standard deviation of the noise uncertainty. This value was
then  multiplied  by  3  to  provide  an  estimate  of  the  $3\sigma$
upper-limit. We  also estimated the  $3\sigma$ `bootstrap upper-limit'
by            multiplying           the           bootstap-uncertainty
$\Delta_{\mathrm{boot}}(\snxs)$ by 3.

The distributions of the  $\snxs$ values for AKN~564 and MCG$-$6-30-15
are quite asymmetric,  with each having an extended  tail towards high
values of $\snxs$. However, the distributions of log~$\snxs$ look much
more symmetric.  This  is not surprising as it is  well known that the
logarithmic transformation of a  random variable with an extended tail
in  its   distribution  brings   that  distribution  much   closer  to
`normality'  \citep[e.g.,][]{pl93}.  Ideally, then,  we would  like to
estimate log~$\snxs$ from each segment  and then determine the mean of
log~$\snxs$ for each object.  Unfortunately, we cannot use this method
because $\snxs$  is negative for  some light curve segments.   We did,
however, determine the logarithm of the mean $\snxs$, which brings the
distribution  of  the  mean  $\snxs$  closer to  normality.   We  also
transformed the uncertainties $\Delta_{\mathrm{tot}}(\snxs)$ to be the
uncertainty in the logarithm of the mean $\snxs$.

The mean $\snxs$ values, uncertainties  and upper limits are listed in
Table~\ref{tab:geninfo}.   The   column  listing  $\snxs$   gives  the
uncertainty  and $3\sigma$  upper limit  as determined  from  only the
bootstrap simulations. The uncertainties and upper-limits given in the
column with log~$\snxs$ include also the noise uncertainty.

\subsection{The variance--mass relation}

The relationship  between log~$\snxs$  and log~$\mbh$ is  presented in
Fig.~\ref{fig:collate_full_log_plawfits_2}. It is  clear that there is
a  strong  anti-correlation  between  the  two  quantities.   This  is
confirmed  using  both  a  Spearman rank-order  correlation  test  and
Kendall's  $\tau$,  both of  which  show  the  anti-correlation to  be
significant  with $>$99.99~per~cent confidence.   The upper  limits to
the variance in  the case where no variability  is detected, which are
shown          in         the          upper          panel         of
Fig.~\ref{fig:collate_full_log_plawfits_2},  are  generally above  the
measured values (for  a given mass).  This means  they are unlikely to
significantly  affect any  model fitting  and  we ignore  them in  the
analysis below.

While there is a strong general  trend for objects with higher mass to
be  less  variable,  there  is  clearly  substantial  scatter  in  the
variance--mass relationship. As we have, for the first time, presented
realistic  estimates  of  the  uncertainties  on  $\snxs$  we  can  be
confident that this scatter is not owing only to these uncertainties.

There  is also  evidence from  the  plot--albeit based  solely on  the
lowest mass object, NGC  4395--that the variance--mass relationship is
non-linear. This  is expected in the  presence of breaks  in the power
spectrum  \citep[e.g.,][]{p04},  as  we  now  show  by  modelling  the
variance--mass  relationship using  both  simple parametrizations  and
with a specific power-spectral form.

%%%%%%%%%%%%%%%%%%%%%%%%%%%%%%%%%%%%%%%%%%%%%%%%%%%%%%%%%%%%%%%%%%%%%%%%%%%
%%%%%%%%%%%%%%%%%%%%%%%%%%%%%%%%%%%%%%%%%%%%%%%%%%%%%%%%%%%%%%%%%%%%%%%%%%%

%%%%%%%%%%%%%%%%%%%%%%%%%%%%%%%%%%%%%%%%%%%%%%%%%%%%%%%%%%%%%%%%%%%%%%%%%%%
%%%%%%%%%%%%%%%%%%%%%%%%%%%%%%%%%%%%%%%%%%%%%%%%%%%%%%%%%%%%%%%%%%%%%%%%%%%

\section{Modelling the relationship between excess variance and mass}

Having obtained  the mean $\snxs$ for  each object, we  then wished to
model the relation between $\snxs$ and $\mbh$. All fits were performed
on log~$\mbh$ and log~$\snxs$.  We fitted the data using both a simple
parametrization and also with a  model that assumes the existence of a
universal power spectrum.

\subsection{Simple parametrizations}

We initially modelled the data using  a power-law of the form $\snxs =
A  \mbh^{-\gamma}$.  The  index and  normalisation of  the best-fiting
power-law were $\gamma  = 0.570$ and $A =  125$, respectively, and the
reduced chi-squared was $\chi_{\nu}^{2}/\mathrm{DOF} = 8.05/31$.  This
model  is   shown  as   a  dot-dashed  line   in  the  top   panel  of
Fig.~\ref{fig:collate_full_log_plawfits_2}.     We   do    not   quote
uncertainties  in  the   best-fitting  parameter  values  because  the
$\chi_{\nu}^{2}$ is formally unsatisfactory.

We then used a singly-broken bending power-law defined as:

\begin{equation}
\snxs   =   A   \mbh^{-\gamma_{\mathrm{low}}}   \left[  1   +   \left(
\frac{\mbh}{M_{\bullet   \mathrm{bend}}}\right)^{\gamma_{\mathrm{high}}
- \gamma_{\mathrm{low}}} \right]^{-1}
\end{equation}

\noindent where $A$ is the normalisation factor and the function bends
from    a    power-law    slope    of    $\gamma_{\mathrm{low}}$    to
$\gamma_{\mathrm{high}}$ at the bend mass $M_{\bullet \mathrm{bend}}$.

We  fixed   the  lower  index  to   $\gamma_{\mathrm{low}}=  0$.   The
best-fitting   bend  mass,   normalisation,  and   upper   index  were
$M_{\bullet \mathrm{bend}}  =$ \scinum{5.59}{5}~$\msun$, $A  = 0.144$,
and      $\gamma_{\mathrm{high}}      =      0.836$,      respectively
($\chi_{\nu}^{2}/\mathrm{DOF}  =  5.99/30$).   The  bending  power-law
clearly improves the fit  statistic substantially, but it is difficult
to assess the formal improvement with, e.g., an F-test as the fits are
so poor.

\subsection{Predicting $\snxs$ from a power spectrum model}

Based on  recent power-spectral analyses  of AGN, it is  possible that
the power  spectra of AGN have  the same shape with  the time-scale of
the variations being proportional to black hole mass (see Introduction
and references  therein).  We decided to  investigate this possibility
by  modelling the  relationship between  $\snxs$ and  $\mbh$  with the
assumption of a universal power spectrum. A model estimate can be made
simply  by  integrating  the   continuous  power  spectrum  over  some
frequency range, for example as defined by the length and the time bin
size of the observation \citep[e.g.,][]{p04}.  This, however, neglects
the effects from the sampling pattern of the light curve, specifically
the  fact  that  it  is  binned,  may have  gaps,  and  is  of  finite
duration. For reasons discussed below these effects, particularly that
of the finite duration and  subsequent `red-noise leak', are likely to
be more  important on  the time-scales considered  here than  the much
longer  ones discussed by  \citet{p04}.  We  have taken  an analytical
approach to  determining the model-predicted $\snxs$,  rather than use
simulations    as   is   typical    for   power    spectrum   analysis
\citep[e.g.,][]{ump02}.  Our  approach is preferable  for two reasons.
First, simulations  are far  more computer-intensive, and  second they
rely   on  the   simulation  technique   accurately   reproducing  the
characteristics   of  the   physical  process   giving  rise   to  the
variability.    While  the  technique   described  below   applies  to
calculation   of   model   $\snxs$    values   it   can   be   adapted
straightforwardly to the estimation of discrete model power-spectra.
 
According to Parseval's Theorem, the  variance in a binned light curve
is  equal to the  sum of  the powers  in the  observed \emph{discrete}
power  spectrum  of  that  light  curve.  The  model  power  spectrum,
however,  is  initially defined  in  a  functional  form and  is  thus
\emph{continous}.   We  denote   this  continuous  power  spectrum  as
$P_{\mathrm{M}}(\nu)$.  We  need to  determine how the  discrete power
spectrum  $P_{\mathrm{D}}(\nu)$ is  related  to $P_{\mathrm{M}}(\nu)$.
The  following description  is  appropriate for  evenly sampled  light
curves containing  no gaps  and having an  even number of  bins.  Note
also  that  the model  power  spectrum  $P_{\mathrm{M}}(\nu)$ must  be
defined to be \emph{two-sided} and,  since we are dealing with a noise
process, we refer to the expectation value of each power.

The first effect to consider  is binning. Suppose we have a continuous
process, with  power spectrum $P_{\mathrm{M}}(\nu)$,  and we transform
it into a discrete process by binning the signal over a time period of
$\delta t$. The power spectrum of the observed binned light curve, say
$P_{\mathrm{B}}(\nu)$, is related to the $P_{\mathrm{M}}(\nu)$ through
the following relation:

\begin{equation}
<P_{\mathrm{B}}(\nu)> = B(\nu) P_{\mathrm{M}}(\nu)
\end{equation}

\noindent where the binning function $B(\nu)$ \citep{v89} is given by:

\begin{equation}
B(\nu) =  \left[ \frac{\mathrm{sin}(\pi \nu \delta  t)}{\pi \nu \delta
t}\right]^{2}
\end{equation}

The next  effect to consider is  aliasing. The fact  that the observed
light curve is sampled at discrete intervals means that power can leak
into the power spectrum from above the Nyquist frequency $\nunyq = 1 /
(2\delta   t)$.   The   binned   and  aliased   power  spectrum,   say
$P_{\mathrm{BA}}(\nu)$,  is related  to the  intrinsic  power spectrum
$P_{\mathrm{M}}(\nu)$ through the relation \citep{p89}:

\begin{equation}
<P_{\mathrm{BA}}(\nu)> = \sum_{i=-\infty}^{\infty} <P_{\mathrm{B}}(\nu
+ i/\delta t)>
\end{equation}

\noindent  The  power  in  one  of our  typical  model  power  spectra
decreases sharply with frequency and  the data are binned.  This means
that  only a  relatively small  amount of  power is  aliased  into the
observed frequency range. Accordingly, we found that summing from $i =
-10$ to $i = 10$ was easily sufficient to account for aliasing.  Power
spectra that are either flat  or increase with frequency might require
a larger range in $i$.

The final effect  to account for is red-noise  leak.  This occurs when
variations  exist  at frequencies  lower  than  those  sampled by  the
observed light  curve, as  is the case  for a red-noise  process. This
`leakage' of power from low to  high frequencies can be seen as either
a rising or  falling trend over the duration of  the light curve.  The
power spectrum of the  final light curve, i.e.  $P_{\mathrm{D}}(\nu)$,
is    related    to     the    intrinsic    power    spectrum,    i.e.
$P_{\mathrm{M}}(\nu)$,      by     the     convolution      of     the
$<P_{\mathrm{BA}}(\nu)>$ with the so-called `window function' $W(\nu)$
of the observed light curve. For evenly-sampled light curves, $W(\nu)$
is simply Fejer's kernel \citep[e.g.,][]{p89}:

\begin{equation}
W(\nu)  = \frac{1}{T} \left[  \frac{\mathrm{sin}(\pi \nu  T)}{\pi \nu}
\right]^{2}
\end{equation}

\noindent where $T$  is the duration of the  light curve. We performed
the convolution with the numerical integral:

\begin{equation}
P_{\mathrm{D}}(\nu)      =     2      \sum_{i      =     -Nf/2}^{Nf/2}
<P_{\mathrm{BA}}(i\delta \nu')>~~W(\nu - i\delta \nu') \delta \nu'
\label{eqn:conv}
\end{equation}

\[ (\nu = 1/T, 2/T ,...,\nunyq) \]

\noindent In  the above sum,  $N$ is the  number of bins in  the light
curve and $f$ is a  positive integer. The frequency step $\delta \nu'$
is given by $\delta  \nu' = 1 / (Tf)$. The value  of $f$ must be large
enough so  that the convolution  extends to a low-enough  frequency to
account  for all of  the low-frequency  power. Determining  a suitable
value of $f$ required a  process of trial-and-error.  We performed the
convolution  with  successively higher  values  of  $f$ until  further
increases produced only  a negligible effect. We found  that $f = 500$
was sufficent for all our convolutions.  Note that the introduction of
the factor  2 in Eqn.~\ref{eqn:conv}  means that $P_{\mathrm{D}}(\nu)$
is \emph{single-sided}  and it is defined only  for $N/2$ frequencies.
Also note that, for $i\delta  \nu' = \pm\nunyq$ the term $\delta \nu'$
was replaced by  $\delta \nu' / 2$, to account  for the end-effects in
the numerical integral.

The expected excess variance was then determined by summing the powers
in $P_{\mathrm{D}}(\nu)$:

\begin{equation}
\sigma_{NXS,model}^{2} = \left[ \sum_{i=1}^{N/2-1} P_{\mathrm{D}}(i/T)
\delta \nu \right] + \frac{1}{2} P_{\mathrm{D}}(\nunyq) \delta \nu
\end{equation}

\noindent where $\delta  \nu = 1/T$.  The factor  of $1/2$ is required
for $P_{\mathrm{D}}(\nunyq)$ because  in a double-sided power spectrum
the Nyquist  frequency occurs only  once.  The factor $\delta  \nu$ is
required   because    the   power    is   expressed   in    units   of
fractional-rms-squared per Hz.

Each of our  305 light curve segments has  its own particular duration
and  sampling pattern,  and there  are many  gaps in  the  data train.
Therefore, the window function will  be different for each segment and
will not, in general, be  represented by Fejer's kernel.  However, the
presence  of missing  bins in  the light  curve will  affect  only the
scatter  in  the  $\snxs$  measurements,  with the  mean  value  being
unaffected.  Moreover, we have taken  care to use light curve segments
of  similar  durations.   Therefore,  we  were able  to  simplify  the
modelling procedure by  assuming that our light curves  were all fully
sampled with the same  number of bins.  We used $N =  148$, as this is
the  even  number-of-bins closest  to  the  mean  segment duration  of
38143.5~s.   Having  made this  simplification,  we  were required  to
determine  only a  single value  of $\sigma_{NXS,model}^{2}$  for each
object  (for a  certain model  power spectrum),  thus speeding  up the
modelling process.

\subsection{A universal power spectrum model}

Motivated by  power-spectral analyses of  AGN \citep[see Introduction,
in   particular][]{mev03},   and   following   the  recent   work   of
\citet[][]{p04},  we hypothesised  a universal  power spectrum  of the
form:

\begin{equation}
P_{\mathrm{M}}(\nu) = A~( \nulfb / \nuhfb )^{-1} (\nu \le \nulfb)
\end{equation}

\begin{equation}
P_{\mathrm{M}}(\nu) = A~( \nu / \nuhfb )^{-1} (\nulfb < \nu < \nuhfb)
\end{equation}

\begin{equation}
P_{\mathrm{M}}(\nu) = A~( \nu / \nuhfb )^{-2} (\nuhfb \le \nu)
\end{equation}

\noindent  where the  normalisation factor  $A$  is the  power at  the
high-frequency break  $\nuhfb$.  The value  of $\nuhfb$ is  assumed to
decrease with black  hole mass, according to the  expression $\nuhfb =
\chfb / \mbh$,  where $\chfb$ is a constant and $\mbh$  is the mass of
the black hole in units of $\msun$. The low-frequency break is related
to  the high-frequency  break by  $\nulfb =  \nuhfb /C_{\mathrm{LFB}}$
where  $\clfb$ is  a  constant.   The normalisation  $A$  varies as  a
function of  $\nuhfb$ as  $A = \psdamp  / \nuhfb$, where  $\psdamp$ is
assumed to be the same for all objects. Using this model, the relation
between  variance  and mass  can  therefore  be  described with  three
parameters: $\chfb$, $\clfb$, and $\psdamp$.

To determine the best-fitting  model, we minimised $\chi^{2}$ for grid
of values of $\chfb$, $\clfb$, and $\psdamp$ values.  We found that we
could  not  constrain the  parameter  $\clfb$.   This  is because  the
low-frequency  break  generally  does  not  fall  within  our  sampled
frequency range.  Therefore,  we fixed this at $\clfb  = 20$.  This is
roughly   the  value  of   $\clfb$  observed   in  the   AGN  NGC~3783
\citep{mev03}    and    in    Cyg~X-1    in   the    low/hard    state
\citep{bh90b,nvw99}.

\begin{figure}

  \rotatebox{270}{\includegraphics[height=1\columnwidth]{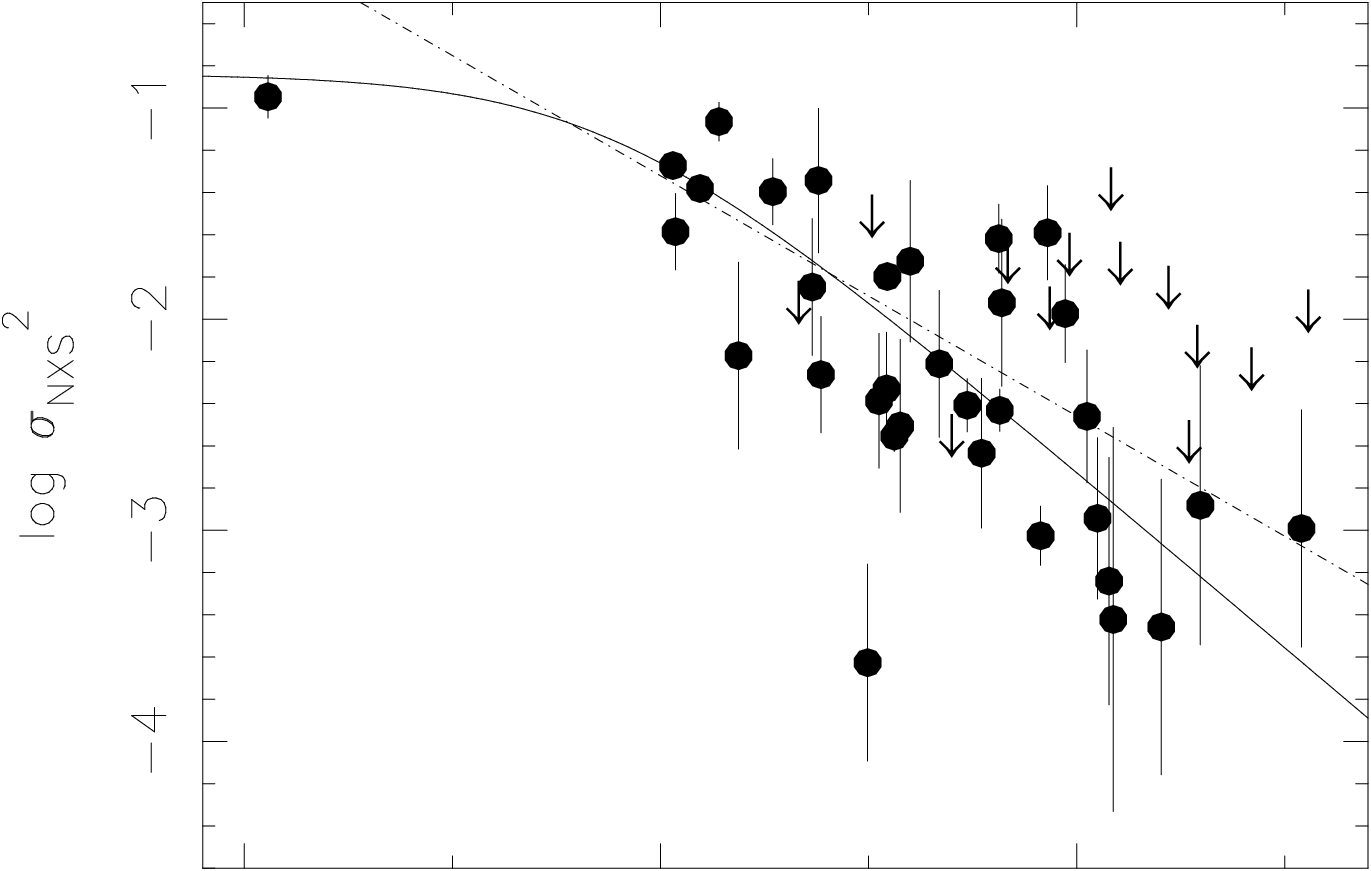}}
  \rotatebox{270}{\includegraphics[height=1\columnwidth]{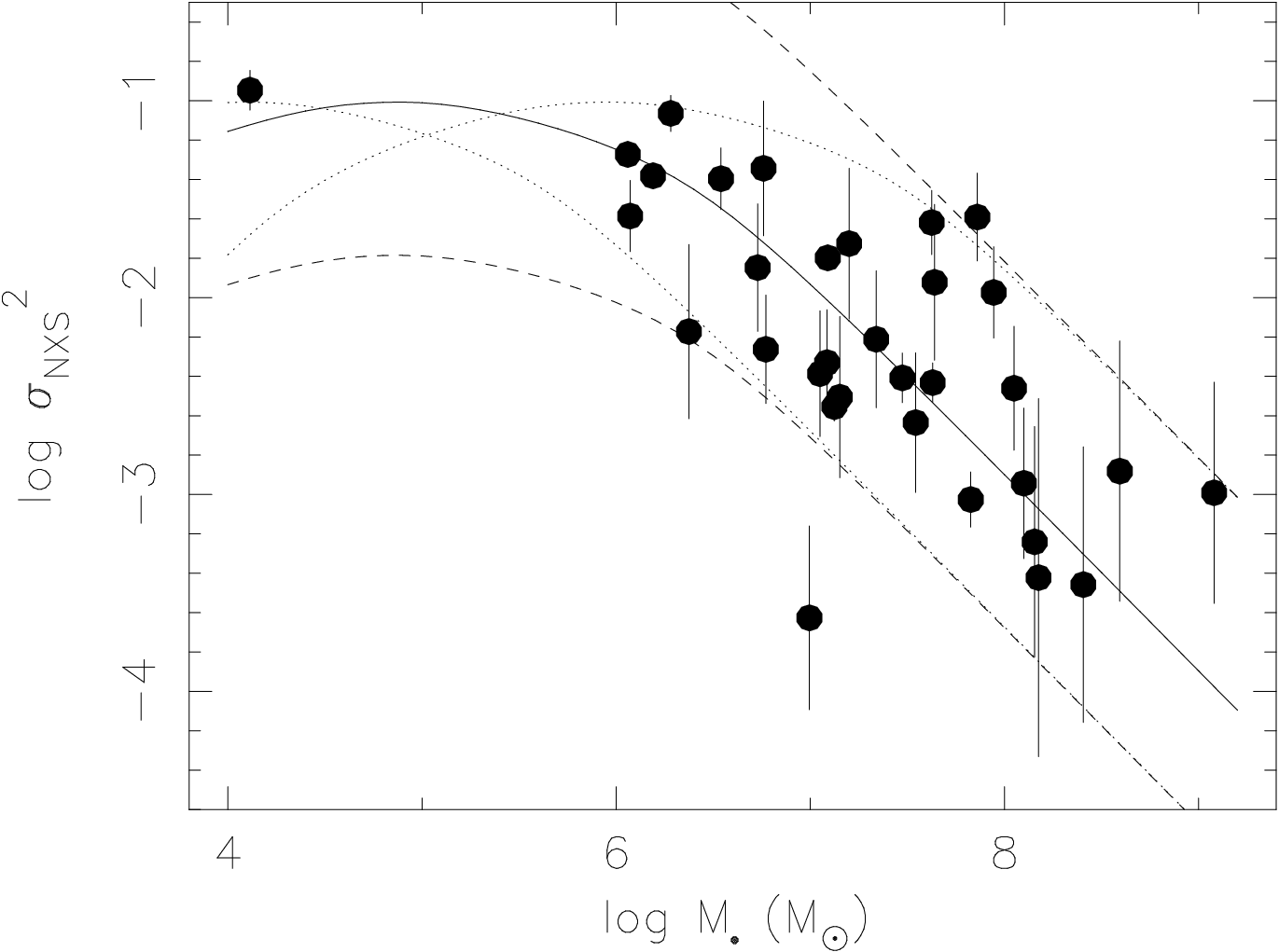}}

\caption{Log of excess variance versus log of black hole mass.  In the
top panel, the dot-dashed and  solid lines show the best-fitting power
law and bending power-law  models, respectively.  In the bottom panel,
the solid line shows best-fitting universal power spectrum model.  The
dotted  and  dashed lines  illustrate  the  effect  of varying  either
$\chfb$  or  $\psdamp$,  respectively  (see text  for  details).   The
$\snxs$ upper limits are, for clarity, shown only in the upper panel.}

\label{fig:collate_full_log_plawfits_2}
\end{figure}

The  best-fitting  values  of  $\chfb$  and  $\psdamp$  are  given  in
Table~\ref{tab:fits}.  This best-fitting  model (for the fit including
all    33    objects)   is    shown    as    the    solid   line    in
Fig.~\ref{fig:collate_full_log_plawfits_2}~(bottom).   The probability
of exceeding the $\chi^{2}_{\nu}$  of the best-fitting universal model
is \scinum{2}{-25}.   This indicates that, while the  model appears to
describe rather  well the overall  trend of decreasing  $\snxs$, there
exists  significant  scatter not  accounted  for  by  the model.   The
residuals $\Delta  \mathrm{log}~\snxs$ from  this model are  listed in
Table~\ref{tab:geninfo}.  We  also fitted  the universal model  to the
data  excluding  various objects.   As  seen in  Table~\ref{tab:fits},
neither the lowest mass object (viz, NGC~4395), nor the 6 objects with
the   largest  number   of   light  curve   segments  (viz,   AKN~564,
MCG$-$6-30-15, TON~S180,  NGC~4151, NGC~3516, NGC~5548),  dominate the
fit.

The  scatter  present  in  the relationship  between  log~$\snxs$  and
log~$\mbh$  can be  explained with  a variation  of either  $\chfb$ or
$\psdamp$  from their  best-fitting  values.  This  is illustrated  in
Fig.~\ref{fig:collate_full_log_plawfits_2}~(bottom).   We find  that a
range  in  $\chfb$  values  between  7.2  and  520  (upper  and  lower
dotted-lines, respectively), or a range in $\psdamp$ between 0.004 and
0.29  (upper and lower  dashed lines,  respectively), can  account for
most of the scatter in the log~$\snxs$ versus log~$\mbh$ relation.

The scatter might also be  owing to a combination of the uncertainties
in  log~$\snxs$  \emph{and}  log~$\mbh$,   the  latter  of  which  are
typically  about  0.5~dex  \citep[e.g.,][]{wu02,pfg04}.  We  performed
simulations to investigate  this possiblity, adopting the best-fitting
relation between  log~$\snxs$ and log~$\mbh$ as our  model.  We needed
first to obtain a  set of 33 model data points to  which we could then
apply scatter in log~$\snxs$ and log~$\mbh$.  To do this, we projected
each of  our 33 observed  data points onto the  best-fitting relation,
minimising  the distance  between the  observed point  and  the model.
(The distance between an observed data point and any particular location on
the model relation was  calculated from the differences in log~$\snxs$
and log~$\mbh$  between the observed  point and the model,  divided by
the corresponding uncertainty in the  observed values.)   Having thus
adopted  a  set  of 33  model  data  points,  we then  performed  1000
simulations.   Each of  these  involved adding  scatter  to the  model
points and then determining the $\chi_{\nu}^{2}$ between the simulated
data points and the model  relation.  We found that 79~per~cent of the
simulations produced a $\chi_{\nu}^{2}$ exceeding that found for the
observed data.   Therefore, the scatter  that we have observed  in the
relation  between log~$\snxs$ and  log~$\mbh$ might  be owing  only to
measurement uncertainties.  If this is  indeed the case, then we would
expect this scatter to be unrelated to other properties of the objects
in our  sample, and we  investigate this possibility in  the following
Section.

%%%%%%%%%%%%%%%%%%%%%%%%%%%%%%%%%%%%%%%%%%%%%%%%%%%%%%%%%%%%%%%%%%%%%%%%%%%
%%%%%%%%%%%%%%%%%%%%%%%%%%%%%%%%%%%%%%%%%%%%%%%%%%%%%%%%%%%%%%%%%%%%%%%%%%%

\begin{table}

\begin{threeparttable}

\caption{Best-fitting  values  for  fits  using  the  universal  power
spectrum model.}

\begin{tabular}{lccc}
\hline

Excluded objects & $\chfb$ & $\psdamp$ & $\chi_{\nu}^{2}/\mathrm{DOF}$
\\ & (Hz $\msun$) & & \\ (1) & (2) & (3) & (4) \\

\hline

None (all objects & 43 & 0.024 & 6.24/31 \\ are included) \\ \\

NGC~4395 & 53 & 0.021 & 6.30/30 \\ \\

AKN~564, & 55 & 0.033 & 4.30/24 \\

MCG$-$6-30-15, \\

TON~S180, NGC~4151, \\

NGC~3516, NGC~5548 \\

\hline

\end{tabular}

{\small (1)  Objects excluded from  fit. (2) Scaling constant  for the
high-frequency  break $\nuhfb$,  where $\nuhfb  = \chfb  /  \mbh$. (3)
Power-spectral  amplitude  at  $\nuhfb$  in power  $\times$  frequency
space. (4) Reduced chi-squared and degrees-of-freedom for fit.}

\label{tab:fits}

\end{threeparttable}
\end{table}

%%%%%%%%%%%%%%%%%%%%%%%%%%%%%%%%%%%%%%%%%%%%%%%%%%%%%%%%%%%%%%%%%%%%%%%%%%%
%%%%%%%%%%%%%%%%%%%%%%%%%%%%%%%%%%%%%%%%%%%%%%%%%%%%%%%%%%%%%%%%%%%%%%%%%%%

%%%%%%%%%%%%%%%%%%%%%%%%%%%%%%%%%%%%%%%%%%%%%%%%%%%%%%%%%%%%%%%%%%%%%%%%%%%
%%%%%%%%%%%%%%%%%%%%%%%%%%%%%%%%%%%%%%%%%%%%%%%%%%%%%%%%%%%%%%%%%%%%%%%%%%%

%%%%%%%%%%%%%%%%%%%%%%%%%%%%%%%%%%%%%%%%%%%%%%%%%%%%%%%%%%%%%%%%%%%%%%%%%%%
%%%%%%%%%%%%%%%%%%%%%%%%%%%%%%%%%%%%%%%%%%%%%%%%%%%%%%%%%%%%%%%%%%%%%%%%%%%

%%%%%%%%%%%%%%%%%%%%%%%%%%%%%%%%%%%%%%%%%%%%%%%%%%%%%%%%%%%%%%%%%%%%%%%%%%%
%%%%%%%%%%%%%%%%%%%%%%%%%%%%%%%%%%%%%%%%%%%%%%%%%%%%%%%%%%%%%%%%%%%%%%%%%%%

\begin{figure}
\includegraphics[width=1\columnwidth]{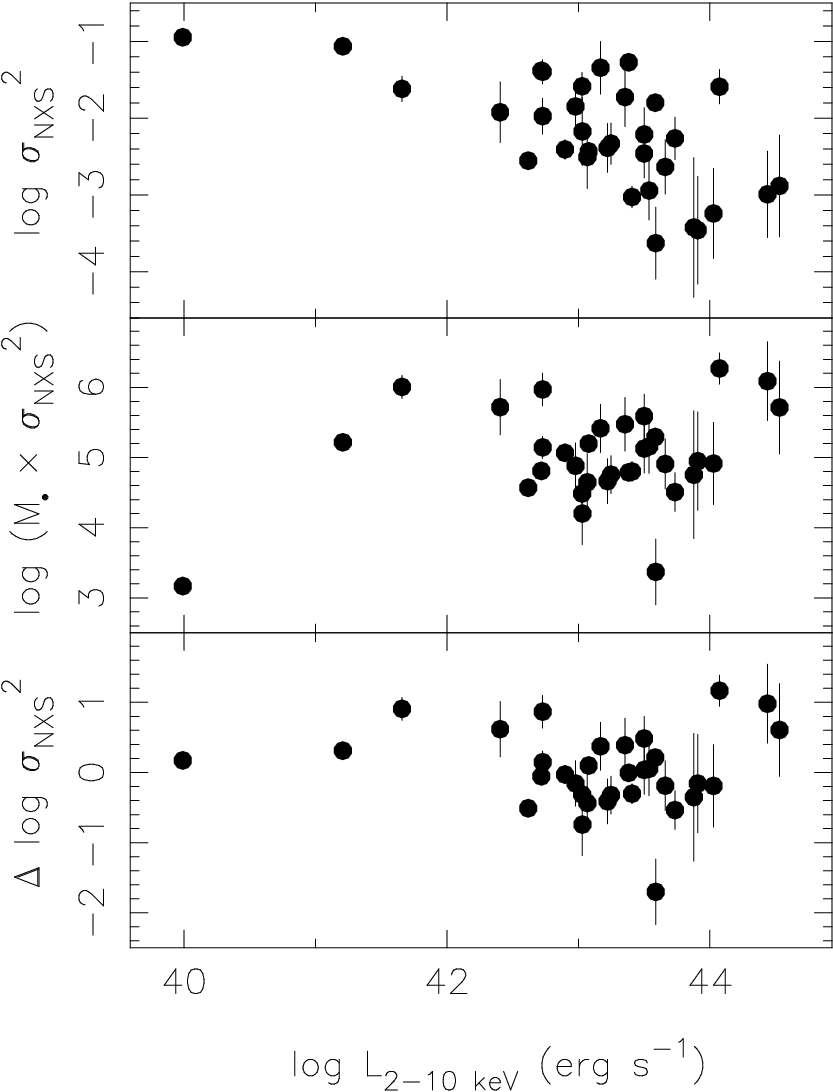}

\caption{Log of  excess variance (top),  log of the product  of excess
variance and  black hole mass (middle), and  excess variance residuals
(bottom), versus log of the 2--10~keV luminosity.}

\label{fig:lum_plot}

\end{figure}

%%%%%%%%%%%%%%%%%%%%%%%%%%%%%%%%%%%%%%%%%%%%%%%%%%%%%%%%%%%%%%%%%%%%%%%%%%%
%%%%%%%%%%%%%%%%%%%%%%%%%%%%%%%%%%%%%%%%%%%%%%%%%%%%%%%%%%%%%%%%%%%%%%%%%%%

%%%%%%%%%%%%%%%%%%%%%%%%%%%%%%%%%%%%%%%%%%%%%%%%%%%%%%%%%%%%%%%%%%%%%%%%%%%
%%%%%%%%%%%%%%%%%%%%%%%%%%%%%%%%%%%%%%%%%%%%%%%%%%%%%%%%%%%%%%%%%%%%%%%%%%%

\begin{figure}
\includegraphics[width=1\columnwidth]{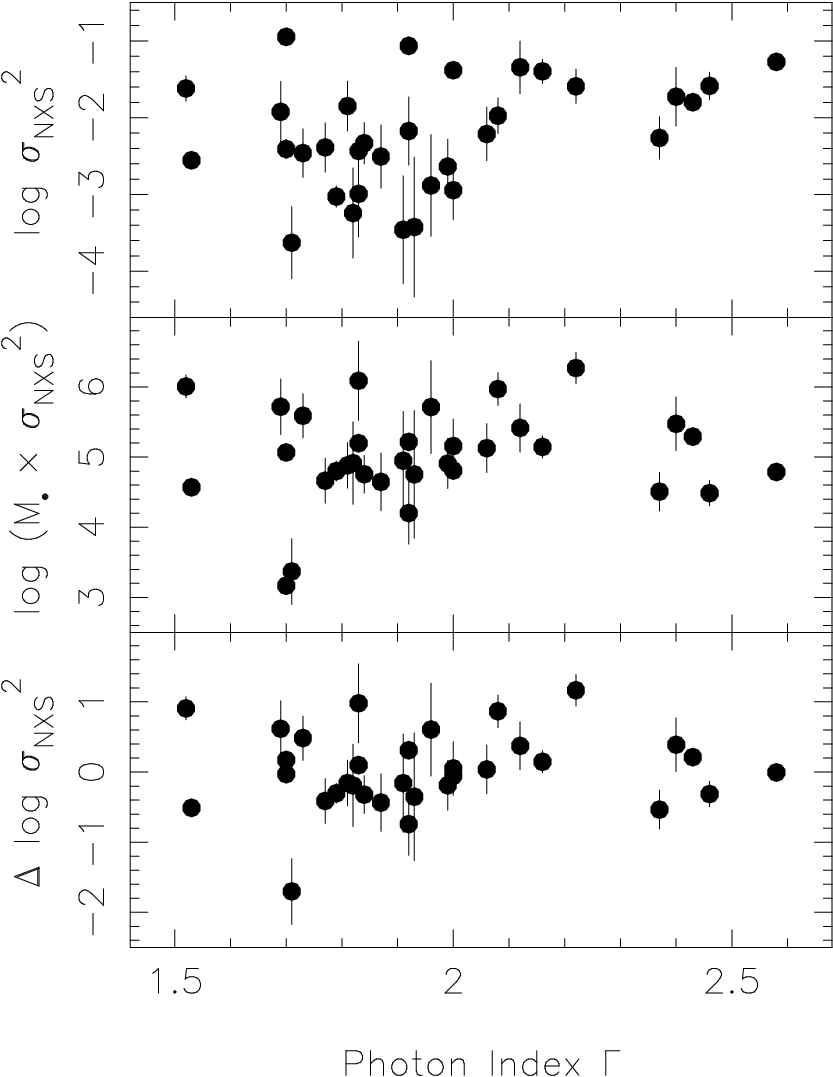}

\caption{Log of  excess variance (top),  log of the product  of excess
variance and  black hole mass (middle), and  excess variance residuals
(bottom), versus the 2--10~keV photon index.}

\label{fig:gamma_plot}

\end{figure}

%%%%%%%%%%%%%%%%%%%%%%%%%%%%%%%%%%%%%%%%%%%%%%%%%%%%%%%%%%%%%%%%%%%%%%%%%%%
%%%%%%%%%%%%%%%%%%%%%%%%%%%%%%%%%%%%%%%%%%%%%%%%%%%%%%%%%%%%%%%%%%%%%%%%%%%

%%%%%%%%%%%%%%%%%%%%%%%%%%%%%%%%%%%%%%%%%%%%%%%%%%%%%%%%%%%%%%%%%%%%%%%%%%%
%%%%%%%%%%%%%%%%%%%%%%%%%%%%%%%%%%%%%%%%%%%%%%%%%%%%%%%%%%%%%%%%%%%%%%%%%%%

\begin{figure}
\includegraphics[width=1\columnwidth]{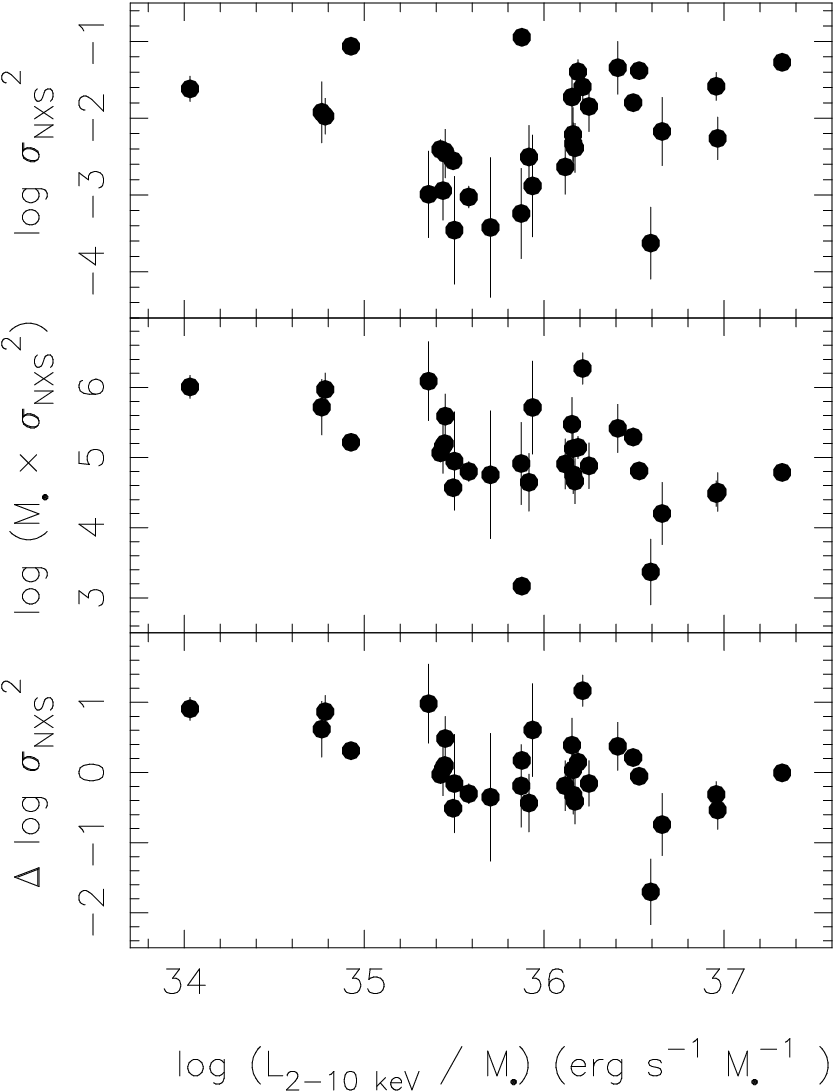}

\caption{Log of  excess variance (top),  log of the product  of excess
variance and  black hole mass (middle), and  excess variance residuals
(bottom),  versus log of  the 2--10~keV  luminosity normalised  by the
black hole mass.}

\label{fig:normlum_plot}

\end{figure}

%%%%%%%%%%%%%%%%%%%%%%%%%%%%%%%%%%%%%%%%%%%%%%%%%%%%%%%%%%%%%%%%%%%%%%%%%%%
%%%%%%%%%%%%%%%%%%%%%%%%%%%%%%%%%%%%%%%%%%%%%%%%%%%%%%%%%%%%%%%%%%%%%%%%%%%

\section{The origin of the scatter in the variance--mass relationship}

Previous studies have revealed an anti-correlation between $\snxs$ and
X-ray  luminosity,  and a  positive  correlation  between $\snxs$  and
photon  index  $\Gamma$ \citep[e.g.,][]{ngm97,tgn99,me01,p04}.   Given
the  strong  dependence between  the  $\snxs$  and  $\mbh$, it  is  of
interest  to see  whether  these correlations  still  exist when  this
primary dependence is removed.  This  should allow us to shed light on
the origin of the scatter in the variance--mass relationship.

In Fig.~\ref{fig:lum_plot}  we plot log $\snxs$,  log $\mbh\snxs$, and
the  residuals  $\Delta   \mathrm{log}~\snxs$  from  the  best-fitting
universal  model, versus  the logarithm  of the  2--10~keV luminosity.
The  quantities log~$\mbh\snxs$  and  $\Delta \mathrm{log}~\snxs$  are
useful  because  they  remove  the  mass-dependence.   Note  that  the
quantity     log     $\mbh\snxs$     is     model-independent.      In
Figs.~\ref{fig:gamma_plot}  and  \ref{fig:normlum_plot}  we  plot  the
variability parameters versus, respectively,  the photon index and the
logarithm  of the 2--10~keV  luminosity normalised  to the  black hole
mass, log~($L_{2-10~\mathrm{keV}}  / \mbh$).   To the extent  that the
X-ray luminosity  is proportional to the bolometric  luminosity, as is
commonly  assumed, the value  log~($L_{2-10~\mathrm{keV}} /  \mbh$) is
proportional  to the ratio  between the  mass-accretion rate  and that
required  to  reach the  Eddington  luminosity  (i.e., the  `Eddington
ratio').  Note  that the correction  factor between the  2--10~keV and
bolometric luminosities is  uncertain, with considerable scatter.  The
Spearman  rank-order correlation coefficient  and Kendall's  $\tau$ of
all 9 relationships are presented in Table~\ref{tab:coeffs}.

As with  previous studies, we  find a very strong  correlation between
log~$\snxs$         and        log~$L_{2-10~\mathrm{keV}}$        (see
Fig.~\ref{fig:lum_plot}).  This correlation  disappears when we remove
the dependence  of $\snxs$ on $\mbh$.   It seems most  likely that the
primary  correlation is  in  fact  with mass,  and  that the  apparent
correlation with log~$L_{2-10~\mathrm{keV}}$ is secondary.

A similar situation is present  when considering the photon index (see
Fig.~\ref{fig:gamma_plot}).     Indeed,   the    correlation   between
log~$\snxs$  and $\Gamma$  is  not  very strong  in  any event,  being
significant  at only  the 96~per~cent  confidence level,  though there
does seem to be an absence of objects having both a steep photon index
and low $\snxs$.  When the  mass dependence is accounted for, however,
no   residual   correlation  remains.    In   the   plot  of   $\Delta
\mathrm{log}~\snxs$ versus $\Gamma$, the steep spectrum objects do not
have  a  systematically higher  $\Delta  \mathrm{log}~\snxs$ than  the
others.

Finally,  we   consider  the  relationship   between  the  variability
properties and  the normalised luminosity log~($L_{2-10~\mathrm{keV}}/
\mbh$)  (see   Fig.~\ref{fig:normlum_plot}).   There  is  considerable
scatter,   and  no   strong  correlation,   between   log~$\snxs$  and
log~($L_{2-10~\mathrm{keV}}/  \mbh$).   Here, however,  we  do find  a
significant  relationship between  log~($L_{2-10~\mathrm{keV}}/ \mbh$)
and both  the mass-normalised excess  variance and the  residuals from
our best-fitting  model.  The  latter correlation is  significant with
$\sim$99~per cent  confidence and, perhaps surprisingly, it  is in the
sense  that objects with  larger values  of normalised  luminosity are
{\it  less}  variable  for  a  given mass.   While  significant,  this
relationship  should be treated  with some  caution.  The  presence of
random scatter in the black  hole mass estimates could possibly induce
such an  anti-correlation.  If  $\mbh$ is underestimated  then $\Delta
\mathrm{log}~\snxs$     will     also     be    underestimated     and
log~($L_{2-10~\mathrm{keV}}  /  \mbh$)   will  be  overestimated.   An
artifical anti-correlation  would certainly be induced  if all objects
had the same value  of log~($L_{2-10~\mathrm{keV}} / \mbh$).  However,
it is  less clear that  this effect could produce  an anti-correlation
between $\Delta  \mathrm{log}~\snxs$ and log~($L_{2-10~\mathrm{keV}} /
\mbh$) in our  data because the normalised luminosities  in our sample
span  3  orders-of-magnitude. We  used  the  simulations described  in
Section~4.3  to test  whether the  observed anti-correlation  could be
owing to the uncertainties in the  black hole masses.  For each of the
1000      simulations,     we      calculated      log~$\snxs$     and
log~($L_{2-10~\mathrm{keV}}  / \mbh$) from  the simulated  data points
and measured Kendall's  $\tau$. We found that, even  with no intrinsic
anti-correlation between  $\snxs$ and $L_{2-10~\mathrm{keV}}  / \mbh$,
57~per~cent  of  the simulations  gave  a  Kendall's  $\tau$ that  was
\emph{more negative} than the observed value of $-$0.31. Therefore, we
cannot  rule-out the  possibility that  the  observed anti-correlation
between $\Delta  \mathrm{log}~\snxs$ and log~($L_{2-10~\mathrm{keV}} /
\mbh$) is an artifact induced  by the presence of uncertainties in the
measurements of black hole mass.

%%%%%%%%%%%%%%%%%%%%%%%%%%%%%%%%%%%%%%%%%%%%%%%%%%%%%%%%%%%%%%%%%%%%%%%%%%%
%%%%%%%%%%%%%%%%%%%%%%%%%%%%%%%%%%%%%%%%%%%%%%%%%%%%%%%%%%%%%%%%%%%%%%%%%%%

\begin{table*}

\begin{threeparttable}

\caption{Correlation coefficients between X-ray variability properties
and the 2--10~keV luminosity, photon index and normalised luminosity.}

\begin{tabular}{llcccc}
\hline

\multicolumn{2}{c}{Observables}   &   \multicolumn{2}{c}{Spearman}   &
\multicolumn{2}{c}{Kendall} \\ & & Coeff. & Sig. (per~cent) & Coeff. &
Sig. (per~cent) \\ (1) & (2) & (3) & (4) & (5) & (6) \\

\hline

log $L_{2-10~\mathrm{keV}}$ & log $\snxs$  & $-0.61$ & 99.98 & $-0.43$
& 99.96 \\

& log $\mbh\snxs$ & 0.13 & 53 & 0.10 & 59 \\

& $\Delta \mathrm{log}~\snxs$ & $-0.06$ & 25 & $-0.04$ & 26 \\

\\

$\Gamma$ & log $\snxs$ & 0.36 & 96 & 0.25 & 96 \\

& log $\mbh\snxs$ & 0.10 & 43 & 0.10 & 56 \\

& $\Delta \mathrm{log}~\snxs$ & 0.11 & 44 & 0.09 & 53 \\

\\

log ($L_{2-10~\mathrm{keV}} / \mbh$) & log  $\snxs$ & 0.29 & 89 & 0.19
& 89 \\

& log $\mbh\snxs$ & $-$0.50 & 99.7 & $-$0.36 & 99.7 \\

& $\Delta \mathrm{log}~\snxs$ & $-$0.44 & 99.0 & $-$0.31 & 98.8 \\

\hline

\end{tabular}

{\small  (1)  X-ray spectral  property  on  the  abscissa.  (2)  X-ray
variability  property  on   the  ordinate.   (3)  Spearman  rank-order
correlation  coefficient.   (4)   Significance  of  correlation.   (5)
Kendall's $\tau$. (6) Significance of correlation.}

\label{tab:coeffs}

\end{threeparttable}
\end{table*}

%%%%%%%%%%%%%%%%%%%%%%%%%%%%%%%%%%%%%%%%%%%%%%%%%%%%%%%%%%%%%%%%%%%%%%%%%%%
%%%%%%%%%%%%%%%%%%%%%%%%%%%%%%%%%%%%%%%%%%%%%%%%%%%%%%%%%%%%%%%%%%%%%%%%%%%

\section{Discussion}

\subsection{Summary of results}

We  have  investigated  the  relationship  between  normalised  excess
variance and black hole mass for  a sample of 46 radio-quiet AGNs.  We
restricted  our light curves  to have  durations between  $\sim$30 and
40~ks  (rest frame), allowing  us to  probe nearly  the same  range of
time-scales for all objects. There  were 32 objects in our sample that
had more than  1 light curve segment. For these  objects, we were able
to  determine the  mean  $\snxs$, decreasing  the  uncertainty in  the
measurements.  Moreover, for 6 objects,  there were more than 15 light
curve segments available.  An  examination of the distributions of the
individual $\snxs$ values  for these 6 objects allowed  us to estimate
the uncertainties in the mean  $\snxs$ for every object in our sample.
These  uncertainties  incorporate  the  effects  of  both  measurement
uncertainties and the stochastic nature of the variability.  Of the 46
objects in our sample, 33 were found to be variable.  As with previous
studies  using   \emph{ASCA}  \citep{ly01,bz03,me04}  and  \emph{RXTE}
\citep{p04,me04} data, we found a significant anti-correlation between
$\snxs$ and $\mbh$.

We initially  fitted the relationship between $\snxs$  and $\mbh$ with
both a  power-law and bending  power-law.  Neither of these  fits were
formally   satisfactory,  however   the  bending   power-law   was  an
improvement over the unbroken power-law.

We  also fitted the  data with  a universal  power spectrum  model. We
determined  the expected  $\snxs$  from  the model  as  a function  of
$\mbh$, accounting for the effects of binning, aliasing, and red-noise
leak    in   the    observed   light    curves.     The   best-fitting
high-frequency-break$\times$mass   scaling-coefficent  was   $\chfb  =
43~\hzmdot$, and the best-fitting amplitude was $\psdamp = 0.024$.  In
his study using \emph{RXTE} data, \citet{p04} found values of $\chfb =
17$ and $\psdamp = 0.017$  ($\chfb = 340$ for NGC~4051).  \citet{me04}
studied the  variability of Seyfert~1 galaxies  on various time-scales
and found  that, on average,  the variability time-scale  followed the
relation   $T_{\mathrm{b}}  =  \mbh   /  10^{6.7}$~days.    Using  our
parametrization,  this corresponds  to a  scaling factor  of  $\chfb =
58~\hzmdot$.

In  general,  the  mass-variance  anti-correlation  can  therefore  be
understood  very simply by  assuming that  all size-scales  scale with
mass, and  hence so do  all characteristic time-scales (such  as those
represented  by  the  break  frequencies).  Our  analysis  furthermore
supports the idea  that the average, or typical  power spectrum of AGN
resembles  the  `universal'   power  spectrum  discussed  above.   The
best-fitting  universal  model  was  not satisfactory,  however,  with
$\chi_{\nu}^{2}/\mathrm{DOF}  =  6.30/31$,   indicating  that,  for  a
certain  $\mbh$,  there  exists  significant scatter  in  the  $\snxs$
values.   However, our  simulations showed  that uncertainties  in the
mass measurements can account for this scatter.

\subsection{The origin of scatter in the variance--mass relation}

Previous work  has suggested  that the excess  variance is  related to
source  properties other  than  mass, such  as  the luminosity,  X-ray
spectral index  and H$\beta$ line  width \citep[e.g.,][]{ngm97,tgn99}.
We have re-investigated some of these relations here.  Consistent with
previous work  using \emph{ASCA} data, we found  a correlation between
log~$\snxs$               and              log~$L_{2-10~\mathrm{keV}}$
\citep[e.g.,][]{ngm97,tgn99,l99}.  The fact that no correlation exists
when the dependence of $\snxs$  on $\mbh$ is removed suggests that the
correlation  between  log~$\snxs$  and log~$L_{2-10~\mathrm{keV}}$  is
largely  a result of  the $\snxs$--$\mbh$  relation.  This  effect has
also been  seen in \emph{RXTE} data  with a time-scale  of about 300~d
\citep{p04}.

We also found  an absence of objects having both  a steep photon index
and low $\snxs$. After accounting for the dependence on mass, however,
we found  no evidence  for a correlation  between excess  variance and
X-ray spectral index. This is perhaps surprising, as previous work has
suggested that  narrow-line Seyfert 1 galaxies--which  have soft X-ray
spectra as a general characteristic \citep[e.g.,][]{bbf96, bme97}--are
more variable  than their broad-line  analogues \citep{tgn99,l99}.  An
effect similar to  that which we have observed  has already been noted
by other  workers using \emph{ASCA} data. \citet{ly01}  found that the
narrow-line Seyfert~1 galaxies in  their sample appeared to follow the
same variance--mass relation as the broad-line objects.  \citet{bz03},
in an expanded study  using the variance measurements of \citet{tgn99}
and  \citet{ly01}, also  found that  the AGN  with $\hbeta$  less than
2000~km~s$^{-1}$ appeared to follow the same relation as those objects
with  broad H$\beta$  emission  lines \citep[see  also the  discussion
in][]{me04}.

We also  found an  anti-correlation between excess  variance residuals
and the normalised  luminosity ($L_{2-10~\mathrm{keV}} / \mbh$), which
we shall  now simply refer to  as the Eddington  ratio $\dot{M}$.  Our
simulations showed that this apparent anti-correlation between $\Delta
\mathrm{log}~\snxs$ and  $\dot{M}$ could be  an artifact owing  to the
uncertainties in the measurements of  the black hole masses.  The fact
that  we did  not find  a \emph{positive}  correlation  between excess
variance and $\dot{M}$, \emph{for a given mass}, is surprising: in the
prevailing  paradigm, NLS1s  generally show  more variability  and are
thought also  to be accreting at high  Eddington ratios \citep{pdo95}.
It  is not  yet  clear, then,  that a  high  value of  $\dot{M}$ is  a
contributing  factor to an  AGN exhibiting  a relatively  large excess
variance.   Further   investigations  in  this   regard  will  benefit
enormously from future improvements in black hole mass measurements.

\subsection{Models for X-ray variability}

In  the  standard  coronal  model,   which  can  be  applied  both  to
stellar-mass  black holes  and AGNs,  seed photons  from  an optically
thick accretion disc are inverse Compton scattered by hot electrons in
an accretion disc corona \citep[e.g.,][]{st80,hm93,cgr01,mr04}.

One class  of models involves the superposition  of individual `shots'
in the  light curve \citep{t72}.  These shots  are possibly associated
with  magnetic  flares  in  the  corona  \citep[e.g.,][and  references
therein]{pf99}.  In the model of \citet{pf99}, there is a distribution
of  shot  time-scales, with  the  value  of  $\nuhfb$ being  inversely
proportional  to the  duration of  the  longest shots.   Also in  that
model,  the variance of  the counting  rate fluctuations  is inversely
proportional to the  mean rate $\lambda$ of the  occurrence of flares.
One can then  assume a basic framework in  which all size-scales (and,
therefore, time-scales) and the  luminosity of the individual shots is
proportional  to  the  black   hole  mass,  accounting  for  the  main
variance--mass relationship.  The  total luminosity is proportional to
$\lambda$, so  for a given black  hole mass the variance  in the light
curve is expected to be inversely proportional to the Eddington ratio.

In the so-called `propagating pertubation' class of models, variations
in the accretion rate occur over  a range of radii from the black hole
\citep[e.g.,][and  references  therein]{l97,cgr01,kcg01,u04}.   Slower
variations  occur  at larger  radii  and  propagate inwards,  coupling
together  with the faster  variations produced  at smaller  radii. The
modulations  in the  accretion rate  propagate to  the  X-ray emission
region and produce variations in the X-ray flux. This type of model is
attractive  because it can  provide an  explanation for  the well-know
`rms--flux'    relation    seen   in    X-ray    binaries   and    AGN
\citep[e.g.,][]{um01,u04,g04}.  The  value of $\nuhfb$  is expected to
be inversely  proportional to  the size of  the X-ray  emission region
because the variations that  originate from within the emission region
are suppressed  \citep{cgr01,u04}. In the model  of \citet{cgr01}, the
low/hard   state  in   Cyg~X-1  occurs   when  the   optically  thick,
geometrically thin  accretion disc is truncated far  from the emission
region. In the  high/soft state, the disc reaches all  the way down to
the emission region  and this leads to the  X-ray variations following
an unbroken  $\alpha = 1$ power-law.   In this model, it  is not fully
specified how the  emission region changes as the  inner radius of the
disc varies. It is clear, however, that the emission region would need
to become smaller as the  disc approaches that region because $\nuhfb$
is higher in the high/soft state than in the low/hard state.

\citet{mpu04} appealed  to the analogy with black  hole X-ray binaries
and speculated  that the location of  the inner edge  of the accretion
disc in AGN is perhaps related to the mass-accretion rate or the black
hole spin.  For  a certain black hole mass,  then, different AGN might
be  regarded as  existing  in  different states,  just  as Cyg~X-1  is
observed in different  states.  In this scenario, we  would expect the
X-ray variability of AGN to be related not only to the black hole mass
but  also the  Eddington ratio  and  photon index.   Objects having  a
relatively high $\dot{M}$ and soft  X-ray spectra would, for a certain
value of  $\mbh$, have  a relatively high  value of $\nuhfb$  (i.e., a
high  value of $\chfb$)  and should,  therefore, exhibit  a relatively
high value of  $\snxs$ for a given range in  time-scales.  We found no
evidence that  the X-ray variability depends on  these properties, and
so  the reality  of this  scenario remains  to be  established.  Note,
however,  that  if an  anti-correlation  existed  between $\chfb$  and
$\psdamp$, then $\chfb$ could  possibly increase without there being a
corresponding increase in $\snxs$.

Discriminating between  various possible scenarios  obviously requires
the use of  power spectral analyses, preferably covering  a wide range
in  source properties.   The challenge,  then, is  to  assemble enough
high-quality power  spectra so that  we can relate  the power-spectral
parameters not  only to $\mbh$ but  also to Eddington  ratio and other
quantities  such as  photon index.   We note,  in particular,  that an
analysis of  the AGN data in the  \emph{XMM-Newton} and \emph{Chandra}
archives, even from relatively  short observations, would be useful in
studying the properties (e.g., power-law slopes) of the variability at
frequencies above the high-frequency break.  A natural starting point,
of course, is  to conduct a rigorous comparison  between the currently
available  power  spectra  \citep[e.g.,][]{ump02,mev03,mpu04} and  the
other  relevent source  properties.  Any  conclusions draw  from these
comparisons  could then be  tested on  a larger  sample of  objects by
using measurements of excess variance.

\section*{Acknowledgments}

The authors  are grateful to  Brad Peterson for kindly  providing some
black hole mass measurements prior  to publication.  We also thank the
anonymous referee for helpful suggestions and comments.  This research
has made use  of the Tartarus (Version 3.0)  database, created by Paul
O'Neill and Kirpal Nandra at  Imperial College London, and Jane Turner
at NASA/GSFC.  Tartarus  is supported by funding from  PPARC, and NASA
grants  NAG5-7385 and  NAG5-7067. PMO  acknowledges  financial support
from PPARC.

\bibliographystyle{mn2e} % if natbib is available

\label{lastpage}

\end{document}